\documentclass[11pt]{article} 
\usepackage[utf8]{inputenc}
\usepackage{simpler-wick}

\usetikzlibrary{decorations.markings}
\tikzset{W->-/.style={decoration={
  markings,
  mark=at position 0.5*\pgfdecoratedpathlength+2pt with
  {\draw[-latex] (-2pt,0pt) -- (1pt,0pt);}},postaction={decorate}},
  W-<-/.style={decoration={
  markings,
  mark=at position 0.5*\pgfdecoratedpathlength with
  {\draw[latex-] (-2pt,0pt) -- (1pt,0pt);}},postaction={decorate}}
  }
\newif\ifWickBelow
\WickBelowfalse
\pgfkeys{
  /simplerwick/.cd,
  arrows/.store in=\LstWickArrows,
  arrows={-,-,-,-,-,-,-,-,-},
  arrows/.initial={-,-,-,-,-,-,-,-,-}, %
  below/.code={\WickBelowtrue},
}

\makeatletter
\def\swick@end#1#2{
  \swick@setfalse@#1
  \tikzexternaldisable
  \begin{tikzpicture}[remember picture, baseline=(swick-close#1.base)]
    \node[use as bounding box, inner sep=0pt, outer sep=0pt] (swick-close#1) {$\displaystyle #2$};
  \end{tikzpicture}
  \tikz[remember picture, overlay]
{
\foreach \W@X[count=\W@C] in \LstWickArrows
{\ifnum\W@C=#1
\xdef\myW@style{\W@X}
\fi}
\ifWickBelow
    \draw[\myW@style] ($(swick-open#1.south) + (0, -3pt)$) 
          -- ($(swick-open#1.base) + (0, -\swick@offset) + #1*(0, -\swick@sep)$) 
          -- ($(swick-close#1.base) + (0, -\swick@offset) + #1*(0, -\swick@sep)$) 
          -- ($(swick-close#1.south) + (0, -3pt)$);
\else
    \draw[\myW@style] ($(swick-open#1.north) + (0, 3pt)$) 
          -- ($(swick-open#1.base) + (0, \swick@offset) + #1*(0, \swick@sep)$) 
          -- ($(swick-close#1.base) + (0, \swick@offset) + #1*(0, \swick@sep)$) 
          -- ($(swick-close#1.north) + (0, 3pt)$);
\fi}
  \tikzexternalenable}
\makeatother

\usepackage{amsmath}
\usepackage{amssymb}
\usepackage{comment}
\usepackage{float}

\usetikzlibrary{decorations.markings}
\usepackage{caption}
\usepackage{xcolor}
\definecolor{dgreen}{rgb}{0,0.5,0}
\definecolor{dpink}{rgb}{1,0.3,0.3}
\definecolor{darkblue}{rgb}{0,0,0.6}
\definecolor{purple}{rgb}{0.4,.2,0.7}
\definecolor{lblue}{rgb}{0.37, 0.60, .95}
\usepackage[margin = 2.5cm]{geometry}
\usepackage{graphicx}
\usepackage[hyperfootnotes = false, colorlinks = true, linkcolor = darkblue, citecolor = darkblue]{hyperref}
\usepackage{subcaption}
\usepackage{transparent}
\usepackage[export]{adjustbox}
\def\X{\mathrm{X}}
\def\O{\mathrm{O}}
\usepackage{cleveref}

\usepackage{fancyhdr}
\parskip=\medskipamount
\newgeometry{margin=2cm}

\def\la{\label}
\def\nref#1{(\ref{#1})}

\usepackage{empheq}
\newcommand*\widefbox[1]{\fbox{\hspace{1em}#1\hspace{1em}}}

\usepackage{amsmath, amssymb, amscd, amsthm, amsfonts}
\usepackage{physics}
\usepackage{graphicx}
\usepackage{multicol}
\usepackage{marvosym}

\newcommand{\bea}{\begin{align}}
\newcommand{\eea}{\end{align}}
\newcommand{\nn}{\nonumber}
\numberwithin{equation}{section}

\newcommand{\be}{\begin{equation}}
\newcommand{\ee}{\end{equation}}
\newcommand{\q}{q}

\def\qb{\bar{Q}}
\def\caln{\mathcal{N}}
\newcommand{\eqn}[1]{\begin{equation}\begin{split} #1 \end{split}\end{equation}}

\renewcommand{\i}{\mathrm{i}}
\newcommand{\lp}{\left (}
\newcommand{\rp}{\right )}
\newcommand{\lb}{\left [}
\newcommand{\rb}{\right ]}
\newcommand{\RA}{\Rightarrow}

\newcommand{\R}{\mathbb{R}}

	\newcommand{\pd}{\partial}
	\newcommand{\inv}{^{-1}}
	\newcommand{\rt}{\sqrt{2}}
	\newcommand{\hf}{\tfrac{1}{2}}

        \def\la{\label}
        \def\nref#1{(\ref{#1})}
	\def\psib{\bar{\psi}}
        \def\Psib{\bar{\Psi}}

        \def\nn{\mathcal{N}}
	\def\cj{\mathcal{J}}
	
	\def\cre{\alpha^{\dagger}}
	\def\nb{{\bar{n}}}

\def\tot{\mathrm{tot}}
\def\max{\mathrm{max}}
 \def\lt{{L}}
 \def\rt{{R}}
 \def\lr{L/R}
 \def\rl{R/L}
\def\TFD{\mathrm{TFD}}

\def\V{{\color{red} \mathsf{V}}} 
\def\vb{{\color{red} \mathsf{\bar{V}} }} 
\def\W{ {\color{blue} \mathsf{W}} }

\begin{document}

\thispagestyle{empty}
\begin{center}
    ~\vspace{5mm}

     {\LARGE \bf Exploring supersymmetric wormholes in $\mathcal{N} = 2$ SYK with chords}
   
   \vspace{0.5in}
     
   {\bf Jan Boruch${}^1$, Henry W. Lin${}^2$, Cynthia Yan${}^2$}

    \vspace{0.5in}

   ~
   \\
   {${}^1$  Department of Physics, University of Warsaw, ul. Pasteura 5, 02-093 Warsaw, Poland}
\,\vspace{0.5cm}\\
{${}^2$ Stanford Institute for Theoretical Physics, Stanford University, Stanford, CA 94305, USA}
                
    \vspace{0.5in}

    \vspace{0.5in}
    
\end{center}

\vspace{0.5in}

\begin{abstract}
A feature the $\mathcal{N}=2$ supersymmetric Sachdev-Ye-Kitaev (SYK) model shares with extremal black holes is an exponentially large number of ground states that preserve supersymmetry.
In fact, the dimension of the ground state subsector is a finite fraction of the total dimension of the SYK Hilbert space.
This fraction has a remarkably simple bulk interpretation as the probability that the zero-temperature wormhole --- a supersymmetric Einstein-Rosen bridge --- has vanishing length. Using chord techniques, we compute the zero-temperature Hartle-Hawking wavefunction; the results reproduce the ground state count obtained from boundary index computations, including non-perturbative corrections. Along the way, we improve the construction \cite{superBerkooz} of the super-chord Hilbert space and show that the transfer matrix of the empty wormhole enjoys an enhanced $\mathcal{N} = 4 $ supersymmetry. 
We also obtain expressions for various two point functions at zero temperature. Finally, we find the expressions for the supercharges acting on more general wormholes with matter and present the superchord algebra. 

\end{abstract}

\vspace{1in}

\pagebreak

\setcounter{tocdepth}{3}

 \tableofcontents

\section{Introduction}

The thermofield double $\ket{\TFD}$ is a fascinating state in holographic theories. For some appropriate range of parameters, $\ket{\TFD}$ is dual to a wormhole \cite{Maldacena:2001kr}, or Einstein-Rosen bridge, that connects two copies of the holographic system, geometrizing their shared entanglement \cite{VanRaamsdonk:2010pw, Maldacena:2013xja}.  %
An interesting question is: what happens in the zero temperature limit $\beta \to \infty$ when the theory has supersymmetric (extremal) black holes? 
It was argued in \cite{Lin:2022rzw, LongPaper} that in certain contexts $\ket{\TFD}$  %
is still described by wormhole of finite length, which can host a variety of states that describe matter propagating inside the wormhole that joins two asymptotic extremal black holes. This suggests an avenue for bringing the technical tools of supersymmetry to bear on the conceptual questions surrounding ER = EPR \cite{VanRaamsdonk:2010pw, Maldacena:2013xja}.

In this work, we explore these supersymmetric wormholes in the concrete setting of the $\mathcal{N} = 2$ double-scaled Sachdev-Ye-Kitaev (DSSYK) model \cite{superBerkooz,Fu:2016vas}. This model was analyzed using chord methods by Berkooz, Brukner, Narovlansky, and Raz \cite{superBerkooz} which we refer to as BBNR\footnote{See also related work on $\caln = 0$ double scaled SYK \cite{Cotler:2016fpe, Berkooz:2018jqr, Berkooz:2018qkz, Pluma:2019pnc, Berkooz:2020uly, Bozejko:1996yv, Lin:2022rbf, newpaper}.}.
Furthermore, the model has an exponentially large number of ground states with exactly zero energy \cite{Fu:2016vas}\footnote{These states are sometimes referred to as BPS states, even though in the JT/SYK context they are annihilated by all the supercharges. In other holographic systems, extremal black holes are BPS states that may only preserve a fraction of the SUSYs.}.
This model is also described by the $\mathcal{N}=2$ super-Schwarzian theory \cite{Stanford:2017thb,Mertens:2017mtv,Lin:2022rzw,LongPaper,Turiaci:2023jfa, Boruch:2023trc}, after taking a further limit\footnote{This limit is the low temperature limit $\beta \to \infty$, taking $\lambda \ll 1$; see \nref{deflambda} for the definition of $\lambda$. In our context we can simply take $\beta \to \infty$ first; more generally, one can take $\beta \to \infty$, $\lambda \to 0$ holding fixed $\lambda \beta$. } known as the ``triple scaling limit.'' The $\nn =2$ Schwarzian theory also describes universal aspects of extremal and near-extremal black holes with a nearly AdS$_2$ throat \cite{Maldacena:2016upp} and an SU$(1,1|1)$ isometry group\footnote{This includes certain extremal and near-extremal black holes in AdS$_5 \cross$S$^5$  \cite{Gutowski:2004ez,Gutowski:2004yv, Chong:2005hr,Cabo-Bizet:2018ehj,Choi:2018hmj,Benini:2018ywd,Boruch:2022tno}, see also the discussion on the closely related $\nn=4$ Schwarzian theory \cite{Heydeman:2020hhw}.}.

An advantage of DSSYK over other holographic quantum systems is that a bulk Hilbert space of sorts (sometimes referred to as the ``chord Hilbert space'') can be derived explicitly from the boundary theory. This Hilbert space describes the states of various Einstein-Rosen wormholes of various lengths and with different matter excitations localized along the wormhole. In the context of $\mathcal{N}=2$ DSSYK, this gives us a new handle to describe supersymmetric wormholes. By studying this chord Hilbert space, we will see concretely how SUSY wormholes encode certain properties of the extremal black hole microstates. 
As an appetizer, we will find a new connection between the classic problem of counting exact zero energy microstates and the geometric properties of the SUSY wormhole:
\eqn{ 
\text{probability(length of SUSY wormhole $=0$)} \; = \; \frac{\dim \mathcal{H}_{\text{extremal}, \, j_R} }{ \dim \mathcal{H}_{j_R} }  \la{superBowlAd}
.}
This formula relates a quantity on the LHS that involves the chord Hilbert space to a quantity on the RHS that involves the microscopic Hilbert space of the boundary theory (the SYK model). In particular, the finite-dimensional Hilbert space $\mathcal{H}_{\text{extremal}, \, j_R}$ in the numerator of \nref{superBowlAd} is the zero energy subspace of the $\nn=2$ SYK model with a fixed $U(1)_{\mathrm{R}}$ charge. In the denominator, the Hilbert space $\mathcal{H}_{j_R}$ contains states of {\it all} energies with fixed charge. (As we will see later, the LHS also implicitly depends on $j_R$; there will be one SUSY wormhole for each possible value of $j_R$.)
We derive this formula around \nref{intro2}. Although we mainly study \nref{superBowlAd} in double scaled SYK, we believe that a similar formula should apply more generally to supersymmetric black holes with a nearly AdS$_2 \times X$ throat, modulo holographic renormalization subtleties; see Section \ref{sec:discussionGen}. 

Note that in the JT gravity regime, the typical length of a SUSY wormhole is large \cite{Lin:2022rzw, LongPaper}, so the quantity on the LHS of \nref{superBowlAd} is a highly ``off-shell'' property of the wormhole.  Indeed, in the gravity regime the fraction on the RHS is quite small\footnote{For finite $p$, $\mathcal{N}=2$ SYK, it is exponentially small in $N$.} so one must be able to compute probabilities to high precision to get a match. We will demonstrate that chord techniques are powerful enough to obtain this probability to high precision, including an infinite series of non-perturbative corrections. These non-perturbative corrections to the ``short wormhole'' probability go beyond the super-Schwarzian/JT gravity approximation \cite{Stanford:2017thb}; however, they are needed to reproduce the microstate count that can be independently inferred from a refined index argument in the microscopic SYK description \cite{Fu:2016vas}, as reviewed in Appendix \ref{app:index}.
Thus the $\mathcal{N}=2$ SYK model is an example of a quantum system where ``simple gravity'' (e.g. the super-Schwarzian) does not reproduce the precise microstate count; one must know some aspects of its UV completion (the super chord theory) in order to obtain certain non-perturbative corrections.

As reviewed in Appendix \ref{app:index}, this gives a leading term for the RHS of \nref{superBowlAd} that agrees with the super-Schwarzian prediction \cite{Stanford:2017thb}. However, there is an infinite series of non-perturbative corrections to the leading answer that are {\it not} reproduced by the super-Schwarzian. We will see in Section \ref{sec:bulkComp} that these are successfully reproduced by the super chord theory. 

The formula \nref{superBowlAd} is a prototype for many of our results. More generally we will find relations where the LHS is a quantity that can be computed using the bulk Hilbert space, e.g., the chord Hilbert space that describes various excitations of the wormhole. On the RHS is a quantity that has a microscopic interpretation, e.g., zero-temperature $n$-pt functions of the SYK model, or equivalently some averaged properties of the one-sided black hole microstates. %

Along the way, we improve various technical aspects of the chord construction of BBNR \cite{superBerkooz}.
First, we clarify the construction of the bulk Hilbert space. The Hilbert space of empty wormholes in $\mathcal{N} = 2$ super JT gravity \cite{Harlow:2018tqv, Lin:2022rzw,LongPaper} is  $L^2(\mathbb{R}) \otimes   L^2(U(1))\otimes\mathbb{C}^4$. The first tensor factor is spanned by wavefunctions $\psi(\tilde \ell),$ where $\tilde \ell$ is the renormalized length of the two-sided wormhole. The second factor is associated to the $U(1)$ Wilson line that stretches between the two sides of the wormhole.
Lastly, we have 2 Dirac fermions (equivalently, two qubits) that are 
the superpartners of the bosonic degrees of freedom.
The Hilbert space of the double-scaled theory will have a similar structure, except that $L^2(\mathbb{R}) \to L^2(\mathbb{Z}_{\ge 0})$, where $L^2(\mathbb{Z}_{\ge 0})$ denotes the Hilbert space of square integrable functions on the non-negative integers. The integer is the chord number, which is interpreted as the discrete bulk length of the wormhole \cite{Lin:2022rbf}.

While the microscopic Hamiltonian has $\mathcal{N} = 2$ SUSY, a new result is that the chord Hamiltonian (or ``transfer matrix'') has enhanced $\mathcal{N} =4$ SUSY when we restrict to acting on states with no matter chords. We will explicitly construct the 4 supercharges. (In the earlier work of BBNR \cite{superBerkooz} the transfer matrix was identified but the enhancement to $\mathcal{N} = 4$ SUSY was not observed. BBNR also found the positive energy eigenfunctions of the transfer matrix, which are scattering states, but the zero energy bound states were not identified.)
We then use these supercharges to find the chord wavefunction corresponding to the supersymmetric wormhole (the $\beta = \infty$ thermofield double). 

This paper is organized as follows: in Section \ref{HilbertSpace}, we discuss the super chord Hilbert space formalism and the action of the SUSY algebra on this Hilbert space. In Section \ref{susyHartleHawking}, we solve for the chord wavefunctions that are annihilated by all the supercharges (the supersymmetric Hartle-Hawking states of the wormhole). %
We use this wavefunction to compute \nref{superBowlAd}. In Section \ref{sec:2pt}, as another application of the Hartle-Hawking wavefunction, we compute the zero temperature 2-pt functions of matter operators. In Section \ref{sec:matter}, we discuss wormholes with matter propagating in the interior. On a technical level, this is relevant for computing more general correlators, such as out-of-time order 4-pt functions, using a higher-pt generalization of \nref{superBowlAd}. On a conceptual level, an interesting feature of supersymmetric wormholes is that despite having zero boundary energy, they can host matter with non-zero ``bulk energy'' in the interior. We derive how the SUSY algebra acts on such wormholes in Section \ref{sec:matter}; furthermore, we show that the SUSY algebra is part of a bigger ``superchord algebra.'' This is a supersymmetric version of the bosonic algebra described in \cite{newpaper}. The superchord algebra should govern the structure of higher-pt functions in the model. We discuss the bulk-to-boundary map in the super SYK model and some conceptual issues about supersymmetric wormholes in Section \ref{discussion}.
Some technical details are presented in the Appendices; for the sake of completeness, we also discuss some aspects of the $\mathcal{N}=1$ SYK model in Appendix \ref{app:n1}. In particular, we found that working out the $\mathcal{N}=1$ superchord algebra was a useful guide to the $\mathcal{N}=2$ case.

\def\bq{\bar{Q}}

\subsection{Review of the model and the chord rules}
The $\mathcal{N}=2$ supersymmetric SYK model \cite{Fu:2016vas} consists of $N$ complex fermions %
\be
\{\psi_i,\bar{\psi}_j\}=\delta_{ij} ,\quad\quad \{\psi_i,\psi_j\}=0  , \la{dirac}
\ee
with interactions governed by the supercharges
\eqn{H = \{Q, \bq \}, \quad Q = \sum_I C_I \Psi_I, \quad \bar Q = \sum_I \bar{C}_I \Psib_I .}
Here $\Psi_I$ is a product of $p$ fermions, $\Psi_I=\psi_{i_1}\cdots\psi_{i_p}$, and $C_I$ are Gaussian complex couplings with mean zero and variance
\be
\langle C_IC^*_{I'}\rangle= 2^{p} \binom{N}{p}^{-1}\mathcal{J}^2\delta_{I,I'}.
\ee
The model enjoys a $U(1)_{\mathrm{R}}$ symmetry, generated by  %
\eqn{ J =   \frac{1}{2p} \sum_{i=1}^N [\psi_i,\psib_i], \quad [J,Q] = Q. \la{U1r}}
We have chosen to normalize\footnote{This differs from BBNR \cite{superBerkooz} conventions by an overall minus sign.} $J$ so that $Q$ has unit charge.
We study the model in the double scaling limit, which is given by
\eqn{N \to \infty, \quad p \to \infty, \quad \lambda \doteq 2p^2/N = \text{fixed}, \quad q  \doteq e^{-\lambda} . \la{deflambda}}
Because in this limit the model becomes exactly solvable, it has been studied extensively in the literature in the context of both nonsupersymmetric \cite{Berkooz:2018jqr,Berkooz:2018qkz} and supersymmetric SYK \cite{superBerkooz}.

In the double scaling limit, correlation functions of $Q, \qb$ and matter operators can be computed using chord techniques \cite{superBerkooz}. The chord method allows one to compute traces of a products of big fermions $\Psi$, by summing over all possible ``chord diagrams''\footnote{In the supersymmetric context, a better terminology would be Qords versus Hamiltonian cHords.} that contribute to the particular trace. Chord diagrams are essentially just a graphical way of representing different Wick contractions, which arise from doing the disorder average over the random couplings. 
Each diagram then contributes a factor that can be read of from the diagram. An example of a single contribution is 
\begin{align}
\label{exampleChordDiagram}
\tr (Q \qb)^3 \V \qb  Q \vb    \hspace{0.5cm} \supset \hspace{0.5cm} 
\begin{tikzpicture}[scale=0.7, baseline={([yshift=0cm]current bounding box.center)}]
	\draw[thick]  (0,0) circle (1.6);
\begin{scope}[thick, decoration={markings, mark=at position 0.2 with {\arrow{>}}}]%
        \draw[postaction=decorate, red] (0.5,-1.5) -- (-1.4,-0.75) ;
        \draw[postaction=decorate]  (0,1.6) node[above] {$Q$} -- (0,-1.6) node[below] {$\bar Q$};
	\draw[postaction=decorate] (1.55025987,-0.39584) node[right] {$Q$} --(-1.05,1.2) node[above left] {$\bar Q$};
	\draw[postaction=decorate] (-1.58,-0.2) node[left] {$Q$} -- (1.13,-1.13) node[right] {$\bar Q$};
	\draw[postaction=decorate] (-1.13,-1.13) node[below left] {$Q$} -- (0.93069294,1.30146481) node[above right] {$\bar Q$};
 \end{scope}
	\draw[fill,black] (-1.58,-0.2) circle (0.1);
	\draw[fill,black] (-1.05,1.2) circle (0.1);
	\draw[fill,black] (1.55025987,-0.39584) circle (0.1);
	\draw[fill,black] (0.93069294,1.30146481) circle (0.1);
	\draw[fill,black] (-1.13,-1.13) circle (0.1);
	\draw[fill,black] (1.13,-1.13) circle (0.1);
	\draw[fill,black] (1.13,-1.13) circle (0.1);
	\draw[fill] (0,-1.6) circle (0.1);
	\draw[fill] (0,1.6) circle (0.1);
        \draw[fill, red] (-1.4,-0.75) circle (0.1) node [left] {$\vb$};
        \draw[fill, red] (0.5,-1.5) circle (0.1) node [below right] {$\V$}; 
\end{tikzpicture} \hspace{0.5cm} =  \hspace{0.5cm} -q^{2+\Delta_\O} %
.
\end{align}
Unlike in the bosonic case, the chords now come with arrows, where the arrows emanate from $Q$ and allow us to keep track of the $U(1)_{\mathrm{R}}$ charge.
Note that without the red chord, the above diagram would vanish since $Q^2 = \qb^2 = 0$. The red chord represents the insertion of some ``matter'' operators $\V$.
Here $\V$ is a product of $p'$ fermions, $\V = K_I \Psi_I, \vb = \bar{K}_I \Psib_I$. 
Note that $\V$ is a chiral or BPS operator since $[Q, \V] = 0$ if $\V$ is bosonic and $\{Q,\V\} =0$ if $\V$ is fermionic.
It will be convenient to introduce $\Delta_\O = p'/p$.

More generally, if $\W \sim \psi_{i_1} \cdots \psi_{i_{p_\O}} \psib_{j_1} \cdots \psib_{j_{p_\X}} $ is a product of $p_\O$ fermions and $p_\X$ anti-fermions, we define $\Delta_\X = p_\X/p $ and $\Delta_\O = p_\O/p$. The $U(1)_{\mathrm{R}}$ charge of such an operator is $\Delta_\O - \Delta_\X$; later, we will see that the conformal dimension\footnote{In general we expect the correlators to look conformally invariant at low temperature and $\lambda \to 0$. For more general temperatures, the correlators should look conformal on a suitably defined fake circle, see \cite{newpaper}.} is 
\eqn{\text{conformal dimension} = \Delta = \hf (\Delta_\X + \Delta_\O) }

Before introducing the general rule for assigning weights to different chord diagrams, let us consider a few simple cases. Consider the correlator
\def\Psib{\Bar{\Psi}}
\eqn{\tr (\Psi_I \Psi_J \Psib_I \Psib_J) = (-1)  e^{-\lambda/2} \la{enemyEx} 
.
}
The only non-zero contributions to this trace occur when $I$ and $J$ do not share any indices in common. If $I,J$ shared any indices, we could use the relation $\psi_i^2 = 0$ and the trace would vanish. We refer to such index pairs  as {\it enemies}. The probability $P(s)$ that some number of indices $s$ are shared between $I$ and $J$ is a Poisson distribution 
\eqn{P(s) = \frac{(\lambda/2)^s}{s!} e^{-\lambda/2} \la{poisson}.}
For enemy configurations, we need $s=0$, which explains the last factor in the RHS \nref{enemyEx}. Furthermore, in \nref{enemyEx} we need to anti-commute the $\Psib_I$ through the $\Psi_J$ before applying the commutation relations. This explains the $(-1)$. %

Let's consider another example where the indices are now in a ``friendly'' configuration:
\eqn{\tr (\Psi_I \Psib_I \Psi_J \Psib_J) =    e^{\lambda/2} \la{friendEx} .}
Unlike the enemy configuration \nref{enemyEx}, here $I, J$ can share indices. The number of indices in common determines the value of the correlator. For example,
\eqn{\tr \psi_1 \psib_1 \psi_1 \psib_1 = 2 \tr \psi_1 \psib_1 \psi_2 \psib_2  .}
We can compare a correlator where $s$ indices are shared $|I\cap J|=s$ to a correlator where $|I\cap J|=0$.
The former will be bigger than the latter by a multiplicative factor of $ 2^s$. Summing over $s$ weighted by \nref{poisson} gives the factor in \nref{friendEx}.

In total, when considering a general Wick contraction contributing to a given moment, there are 4 different possible friend configurations, and 8 enemy configurations. All possible friends and enemies pairs are summarized below: %
\def\Compactcdots{\mathinner{\cdotp\mkern-2mu\cdotp\mkern-2mu\cdotp}}
\begin{subequations}
\begin{align}
&\text{friends:} \qquad \,\, q^{-1/2} &&\wick[arrows={-,W->-,W-<-,-,-}]{ \c2{Q} \Compactcdots \c2{\qb}\Compactcdots   \c2{Q} \Compactcdots \c2{\qb} } , \quad   &&\wick[arrows={-,W->-,-,W-<-,-}]{ \c4{Q} \Compactcdots \c2{\qb} \Compactcdots \c2{Q} \Compactcdots \c4{\qb} } \la{friends} ,\\
& \hspace{2.15cm} q^{-1/2} &&\wick[arrows={-,W-<-}]{ \c2{\qb} \Compactcdots \c2{Q}\Compactcdots   \c2{\qb} \Compactcdots \c2{Q} } , \quad   &&\wick[arrows={-,W-<-,-,W->-,-}]{ \c4{\qb} \Compactcdots \c2{Q} \Compactcdots \c2{\qb} \Compactcdots \c4{Q} } \la{friends2} ,
\end{align}
\end{subequations}
\begin{subequations}
\begin{align}
&\text{enemies:} \qquad q^{+1/2} &&\wick[arrows={-,W->-,W-<-,-,-}]{ \c2{Q} \Compactcdots \c2{\qb} } \Compactcdots \wick[arrows={-,W-<-}]{\c2{\qb} \Compactcdots \c2{Q} } ,  \quad  &&  \wick[arrows={-,W->-,-,W->-,-}]{ \c4{Q} \Compactcdots \c2{Q}  \Compactcdots \c2{\qb} \Compactcdots \c4{\qb} } , \la{enemies1} \\
&\hspace{2.15cm} q^{+1/2} &&\wick[arrows={-,W-<-,}]{ \c2{\qb} \Compactcdots \c2{Q} } \Compactcdots \wick[arrows={-,W->-}]{\c2{Q} \Compactcdots \c2{\qb} } ,  \quad  &&  \wick[arrows={-,W-<-,-,W-<-,-}]{ \c4{\qb} \Compactcdots \c2{\qb}  \Compactcdots \c2{Q} \Compactcdots \c4{Q} } , \la{enemies2} \\
& \hspace{1.8cm} -q^{+1/2} &&\wick[arrows={-,W-<-,-,W->-,-}]{ \c4{Q} \Compactcdots \c2{\qb}  \Compactcdots \c4{\qb} \Compactcdots \c2{Q} } , \quad  && \wick[arrows={-,W->-,-,W->-,-}]{ \c4{Q} \Compactcdots \c2{Q}  \Compactcdots \c4{\qb} \Compactcdots \c2{\qb} }\la{enemies3} ,\\
& \hspace{1.8cm} -q^{+1/2} &&\wick[arrows={-,W->-,-,W-<-,-}]{ \c4{\qb} \Compactcdots \c2{Q}  \Compactcdots \c4{Q} \Compactcdots \c2{\qb} } , \quad  && \wick[arrows={-,W-<-,-,W-<-,-}]{ \c4{\qb} \Compactcdots \c2{\qb}  \Compactcdots \c4{Q} \Compactcdots \c2{Q} }\la{enemies4}
.
\end{align}
\end{subequations}

Every pair of chords of the form \nref{friends} gets a factor $q^{-1/2}$, whereas \nref{enemies1} gives $q^{+1/2}$ and \nref{enemies2} gives $-q^{+1/2}$.
The basic rule of calculating the moments\footnote{There is a factor of $2^{-k}$ that we have omitted in \cite{superBerkooz} by a different choice of normalization.} is 
\eqn{
  (-1)^{\# \text{intersections}} \q^{(\# \text{enemies}-\# \text{friends}) /2}.\la{chordRules}
}
To generalize to the case with chiral matter insertions $\V$, one should weight all pairs of red/black chords by a similar factor but with $q \to q^\Delta$. Red/red pairs of chords would get a factor $q \to q^{\Delta^2}$.

\section{The super chord Hilbert space \la{HilbertSpace}}

\subsection{The two-sided Hilbert space}

In this section, we construct the $\caln=2$ chord Hilbert space. While much of this material is a review of BBNR \cite{superBerkooz}, we will clarify and improve certain aspects of the construction by emphasizing the two-sided nature of the Hilbert space (along the lines of \cite{Lin:2022rzw}) and dealing with the $U(1)_\mathrm{R}$ sector more systematically. In particular we will find that the chord transfer matrix of \cite{superBerkooz}, which may be viewed as the bulk Hamiltonian, enjoys $\caln=4$ supersymmetry (and not just $\caln=2$ discussed in \cite{superBerkooz} ). This parallels the super JT discussion of \cite{LongPaper}. %

The Hilbert space with no matter insertions is generated by acting on the maximally entangled state $\ket{\Omega}$ %
with the operators $ \{ Q_\lt, \bar{Q}_\lt, Q_\rt, \bar{Q}_\rt \}$ as well as the charge generators $J_L$ and $J_R$. %
The starting point of the Hilbert space construction is to view the partition function as an overlap between maximally entangled 2-sided states $\ket{\Omega}$:
\eqn{Z(\mu, \beta) = \tr \lp e^{-\beta H - \mu J}\rp  =\frac{ \bra{\Omega} e^{-\beta H_R - \mu J_R} \ket{\Omega}}{\braket{\Omega}}  \la{Zintro} } 
where the Type II$_1$ normalized trace $\tr 1 = 1$. Equivalently, we could consider integrating over $\mu$, which would project into a fixed $U(1)_{\mathrm{R}}$ charge sector:
\eqn{\label{micro} \frac{Z({j_R}, \beta)}{Z(j_R,0)} = \frac{\tr \lp \Pi_{j_R} e^{-\beta H} \rp}{\tr \Pi_{j_R} } =  \frac{\tr_{j_R} \lp e^{-\beta H}\rp}{\tr_{j_R} (\mathbf{1})  } = \frac{\bra{\Omega,j_R} e^{-\beta H_R} \ket{\Omega,j_R}}{ \braket{\Omega,j_R}} .}
In the microcanonical ensemble $Z(j_R,\beta)$ we only sum over SYK states with charge $J = j_R$.
On the RHS, we have converted to the 2-sided notation. Here $\ket{\Omega} = \sum_j \ket{\Omega,j}$ is the maximally entangled state in the full Hilbert space, whereas $\ket{\Omega,j}$ is the maximally entangled state in a given charge sector $j$.

Now the LHS of \nref{Zintro} can be computed by summing over chord diagrams, which determines all correlators of the supercharges, etc., acting on $\ket{\Omega}$. Taking inspiration from the Gelfand–Naimark–Segal (GNS) construction, this should fully determine the Hilbert space. The idea then is to slice open the chord diagrams and define 2-sided states by considering the data that is needed to compute all correlators acting on these states. As a first pass, we can keep track of the chords that cross the slice; the chord Hilbert space consists of two types of chords: those created by $Q$ (denoted by $\O$ or an $\uparrow$) and chords created by $\bq$ (denoted by $\X$ or a $\downarrow$).  
\begin{align}
\label{ChordDiagramHilbert}
\begin{tikzpicture}[scale=0.7, baseline={([yshift=0cm]current bounding box.center)}]
	\draw[thick]  (0,0) circle (1.6);
\begin{scope}[thick, decoration={markings, mark=at position 0.2 with {\arrow{>}}}]	
        \draw[postaction=decorate, red] (-1.4,-0.75) -- (0.5,-1.5);
        \draw[postaction=decorate]  (0,1.6) node[above] {} -- (0,-1.6) node[below] {};
	\draw[postaction=decorate] (1.55025987,-0.39584) node[right] {} --(-1.05,1.2) node[above left] {};
	\draw[postaction=decorate] (-1.58,-0.2) node[left] {} -- (1.13,-1.13) node[right] {};
	\draw[postaction=decorate] (-1.13,-1.13) node[below left] {} -- (0.93069294,1.30146481) node[above right] {};
 \end{scope}
	\draw[fill,black] (-1.58,-0.2) circle (0.1);
	\draw[fill,black] (-1.05,1.2) circle (0.1);
	\draw[fill,black] (1.55025987,-0.39584) circle (0.1);
	\draw[fill,black] (0.93069294,1.30146481) circle (0.1);
	\draw[fill,black] (-1.13,-1.13) circle (0.1);
	\draw[fill,black] (1.13,-1.13) circle (0.1);
	\draw[fill,black] (1.13,-1.13) circle (0.1);
	\draw[fill] (0,-1.6) circle (0.1);
	\draw[fill] (0,1.6) circle (0.1);
        \draw[fill, red] (-1.4,-0.75) circle (0.1);
        \draw[fill, red] (0.5,-1.5) circle (0.1);
        \draw[dashed] (-1.6,0) -- (1.6,0);
        \draw[fill,gray] (-1.6,0) circle (0.1);
	\draw[fill,gray] (1.6,0) circle (0.1);
\end{tikzpicture} \quad  \RA \quad  \ket{ \uparrow \downarrow \uparrow} = \ket{\O\X\O}
\end{align}
In principle, we need to consider a bulk Hilbert space consisting of arbitrary O,X strings. However, \cite{superBerkooz} showed that it is sufficient to restrict to states where we never have two consecutive X's or two consecutive O's. This is closely related to the fact that $Q^2 = \bq^2 = 0$.  Thus there are four kinds of states:

\begin{align}
\label{string} \ket{n,\X\O} &=  |\overbrace{\X\O \X\O  \cdots \X\O}^{n \text{ pairs of } \X\O} \, \rangle, \quad  && \ket{n,\O\X} = |\overbrace{\O\X \O\X  \cdots \O\X}^{n \text{ pairs of } \O\X} \, \rangle
,\\
 \ket{n, \X\X} &=  |\overbrace{\X\O \X\O  \cdots \X\O}^{n \text{ pairs of } \X\O} \X \rangle, \quad  && \ket{n,\O\O} = |\overbrace{\O\X \O\X  \cdots \O\X}^{n \text{ pairs of } \O\X} \O \rangle 
 .
\end{align}

Here $n$ is a non-negative integer. The fermion number $F$ is defined as the total number of chords in the above state (mod 2):
\eqn{(-1)^F = (-1)^{n_\text{tot}} .}
If there are matter chords, one also multiplies $(-1)^{n_\text{tot}}$ by $(-1)$ for each fermionic matter chord.
Hence $(-1)^F \ket{n,\X\O} = \ket{n,\X\O}$  whereas $ (-1)^F \ket{n, \X\X} = - \ket{n, \X\X}$.
A bosonic state in this theory has an even number of chords, whereas a fermionic state has an odd number of chords.

The notation $\ket{n,ab}$ emphasizes that there are 4 possible states (2 qubits) for each value of $n$. This explains the emergence of the bulk fermions in the $\caln = 2$ super-Schwarzian theory. The bosonic length mode of the wormhole in the super Schwarzian (together with the $U(1)_{\mathrm{R}}$ gauge field) comes with fermionic superpartners, consisting of 2 complex fermions, or equivalently 2 qubits, see also Appendix \ref{schMatch}. %

\subsection{Remembering forgotten friends and enemies}
The above discussion omits an important subtlety, which we now explain. 
In the $\mathcal{N}=0$ and $\mathcal{N}=1$ chord Hilbert space, ket vectors are labeled by a string that describes the open chords on the appropriate slice. The chords which close to the Euclidean ``past'' of this slice are ``forgotten.'' In contrast, the chord rules in the $\caln = 2$ case imply that the action of some operator on a ket vector depends on the number of such closed chords (for examples the first diagram in \nref{friends} and \nref{enemies1}). %
The chord Hilbert space is not just labeled by a string of O's and X's; we need more information to account for closed chords. We do so by including an additional factor of $m$ in the label. For example,
\begin{align}
\label{ChordDiagramHilbert2}
\begin{tikzpicture}[scale=0.7, baseline={([yshift=0cm]current bounding box.center)}]
	\draw[thick]  (0,0) circle (1.6);
\begin{scope}[thick, decoration={markings, mark=at position 0.2 with {\arrow{>}}}]	
        \draw[postaction=decorate, red] (-1.4,-0.75) -- (0.5,-1.5);
        \draw[postaction=decorate]  (0,1.6) node[above] {} -- (0,-1.6) node[below] {};
	\draw[postaction=decorate] (1.55025987,-0.39584) node[right] {} --(-1.05,1.2) node[above left] {};
	\draw[postaction=decorate] (-1.58,-0.2) node[left] {} -- (1.13,-1.13) node[right] {};
	\draw[postaction=decorate] (-1.13,-1.13) node[below left] {} -- (0.93069294,1.30146481) node[above right] {};
 \end{scope}
	\draw[fill,black] (-1.58,-0.2) circle (0.1);
	\draw[fill,black] (-1.05,1.2) circle (0.1);
	\draw[fill,black] (1.55025987,-0.39584) circle (0.1);
	\draw[fill,black] (0.93069294,1.30146481) circle (0.1);
	\draw[fill,black] (-1.13,-1.13) circle (0.1);
	\draw[fill,black] (1.13,-1.13) circle (0.1);
	\draw[fill,black] (1.13,-1.13) circle (0.1);
	\draw[fill] (0,-1.6) circle (0.1);
	\draw[fill] (0,1.6) circle (0.1);
        \draw[fill, red] (-1.4,-0.75) circle (0.1);
        \draw[fill, red] (0.5,-1.5) circle (0.1);
        \draw[dashed] (-1.6,0) -- (1.6,0);
        \draw[fill,gray] (-1.6,0) circle (0.1);
	\draw[fill,gray] (1.6,0) circle (0.1);
\end{tikzpicture} \quad  \RA \quad  \ket{\O\X\O, m= -1-\Delta_\O } .
\end{align} 
Here we have defined 
\begin{align}   
m = (\# \text{ left-moving closed chords}) - (\# \text{ right-moving closed chords}) .
\end{align}
Here we are counting the number of right-moving and left-moving chords that are to the Euclidean past of the gray cut (e.g. part of the ket vector.)
In the absence of matter operators, $m$ will be an integer, but $m$ can be shifted by a fractional amount $m \to m \pm \Delta_\O - \Delta_\X$ by the action of a matter chord.

To understand how this extra label helps, consider cutting open a chord diagram (say along the gray slice) and then inserting an extra chord: 
\begin{align}
\label{exampleChordDiagram2}
\begin{tikzpicture}[scale=0.7, baseline={([yshift=0cm]current bounding box.center)}, decoration={markings, mark=at position 0.2 with {\arrow{>}} }]
	\draw[thick]  (0,0) circle (1.6);
\begin{scope}[thick, decoration={markings, mark=at position 0.2 with {\arrow{>}} }]	
        \draw[postaction=decorate, red] (-1.4,-0.75) -- (0.5,-1.5);
        \draw[postaction=decorate]  (0,1.6) -- (-1.13,-1.13)  ;
        \draw[postaction=decorate]  (0,1.6) -- (-1.13,-1.13)  ;
	\draw[postaction=decorate] (1.55025987,-0.39584) --(-1.05,1.2) ;
	\draw[postaction=decorate] (-1.58,-0.2) -- (1.13,-1.13) ;
	\draw[postaction=decorate] (0,-1.6) -- (0.93069294,1.30146481) ;
 \end{scope}
	\draw[fill,black] (-1.58,-0.2) circle (0.1);
	\draw[fill,black] (-1.05,1.2) circle (0.1);
	\draw[fill,black] (1.55025987,-0.39584) circle (0.1);
	\draw[fill,black] (0.93069294,1.30146481) circle (0.1);
	\draw[fill,black] (-1.13,-1.13) circle (0.1);
	\draw[fill,black] (1.13,-1.13) circle (0.1);
	\draw[fill,black] (1.13,-1.13) circle (0.1);
	\draw[fill] (0,-1.6) circle (0.1);
	\draw[fill] (0,1.6) circle (0.1);
        \draw[dashed] (-1.6,0) -- (1.6,0);
        \draw[fill,gray] (-1.6,0) circle (0.1);
	\draw[fill,gray] (1.6,0) circle (0.1);
        \draw[fill, red] (-1.4,-0.75) circle (0.1);
        \draw[fill, red] (0.5,-1.5) circle (0.1);
\end{tikzpicture} 
\quad \RA \quad 
\begin{tikzpicture}[scale=0.7, baseline={([yshift=0cm]current bounding box.center)}]
	\draw[thick]  (0,0) circle (1.6);
\begin{scope}[thick, decoration={markings, mark=at position 0.2 with {\arrow{>}}}]	
        \draw[postaction=decorate, red] (-1.4,-0.75) -- (0.5,-1.5);
        \draw[postaction=decorate]  (0,1.6) -- (-1.13,-1.13) ;
	\draw[postaction=decorate] (1.55025987,-0.39584) --(-1.05,1.2) ;
	\draw[postaction=decorate] (-1.58,-0.2) -- (1.13,-1.13) ;
	\draw[postaction=decorate] (0,-1.6) -- (0.93069294,1.30146481) ; %
 \end{scope}
	\draw[fill,black] (-1.58,-0.2) circle (0.1);
	\draw[fill,black] (-1.05,1.2) circle (0.1);
	\draw[fill,black] (1.55025987,-0.39584) circle (0.1);
	\draw[fill,black] (0.93069294,1.30146481) circle (0.1);
	\draw[fill,black] (-1.13,-1.13) circle (0.1);
	\draw[fill,black] (1.13,-1.13) circle (0.1);
	\draw[fill,black] (1.13,-1.13) circle (0.1);
	\draw[fill] (0,-1.6) circle (0.1);
	\draw[fill] (0,1.6) circle (0.1);
    \draw[fill,white] (-1.7,0.1) rectangle ++(3.35,2.6); 
        \draw[fill, red] (-1.4,-0.75) circle (0.1);
        \draw[fill, red] (0.5,-1.5) circle (0.1);
        \draw[dashed] (-1.6,0) -- (1.6,0);
\end{tikzpicture}
\quad \RA \quad 
\begin{tikzpicture}[scale=0.7, baseline={([yshift=0cm]current bounding box.center)}]
	\draw[thick]  (0,0) circle (1.6);
\begin{scope}[thick, decoration={markings, mark=at position 0.5 with {\arrow{>}}}]	
        \draw[postaction=decorate, red] (-1.4,-0.75) -- (0.5,-1.5);
        \draw[postaction=decorate]  (0,1.6) -- (-1.13,-1.13)  ;
	\draw[postaction=decorate] (1.55025987,-0.39584) --(-1.05,1.2) ;
	\draw[postaction=decorate] (-1.58,-0.2) -- (1.13,-1.13) ;
	\draw[postaction=decorate] (0,-1.6) -- (0.93069294,1.30146481) ; %
        \draw[color=lblue,postaction=decorate]  (1.4,0.8) node[above right] {$Q$} --   (1.6,0.0) node[above right] {$\bq$}  ;
 \end{scope}
	\draw[fill,black] (-1.58,-0.2) circle (0.1);
	\draw[fill,black] (-1.05,1.2) circle (0.1);
	\draw[fill,black] (1.55025987,-0.39584) circle (0.1);
	\draw[fill,black] (0.93069294,1.30146481) circle (0.1);
	\draw[fill,black] (-1.13,-1.13) circle (0.1);
	\draw[fill,black] (1.13,-1.13) circle (0.1);
	\draw[fill,black] (1.13,-1.13) circle (0.1);
	\draw[fill] (0,-1.6) circle (0.1);
	\draw[fill] (0,1.6) circle (0.1);
    \draw[fill,white] (-1.7,0.1) rectangle ++(2.8,2.6);  
        \draw[dashed] (-1.6,0) -- (1.6,0);
        \draw[fill,lblue] (1.4,0.8) circle (0.1);
        \draw[fill,lblue] (1.6,0.0) circle (0.1);
        \draw[fill, red] (-1.4,-0.75) circle (0.1);
        \draw[fill, red] (0.5,-1.5) circle (0.1);
\end{tikzpicture}
\end{align}
According to the chord rules \nref{chordRules}, this {\color{lblue} new chord} gets a factor of 
$q^{(1 + \Delta_\O)/2 } = q^{-m/2}$
from the two closed chords. (If instead we had added a chord with the opposite orientation, friends and enemies would be reversed and we would have $q^{+m/2}$.)
Furthermore, once we insert this {\color{lblue} new chord}, the resulting state is now $\ket{\O\X\O,m= - \Delta_\O}$. So in general, the action of a chord insertion can be written as $\beta q^{-m/2}$ or $\beta^\dag q^{+m/2}$ depending on the orientation of the chord, where $\beta\ket{m} = \ket{m-1}$ and $\beta^\dag \ket{m} = \ket{m+1}$.

\subsection{Action of the supercharges}%

Now we would like to determine the action of $Q_\rt, \qb_\rt, Q_\lt, \qb_\lt$ on our 2-sided Hilbert space.
Let's consider $Q_\rt$ first. Such an operator can add a new chord, or delete an existing one. If $Q_\rt$ deletes an existing chord, it must delete either the left-most or the right-most chord; otherwise, we would have two consecutive $X$'s or $O$'s. 

Together with the results of the previous subsection, this implies that $Q_\rt$ has the form
\begin{align}
Q_\rt &=  (\text{del X}_\lt  \, \tilde A +   \text{del X}_\rt  \, \tilde B)\beta + q^{m/2} \text{add O}_\rt   \\
&= (A \,  \text{del X}_\lt  +   B\, \text{del X}_\rt) \beta   +q^{m/2} \text{add O}_\rt 
\end{align}
The factors of $\beta$ and $q^{m/2}$ perform the forgotten frenemy accounting. As explained after \nref{exampleChordDiagram2}: every time we delete a chord, we increment/decrement $\ket{m} \to \ket{m\pm 1}$, since closing a chord implies that there is a new ``forgotten friend'' or ``forgotten enemy'' added to the list. On the other hand, when we add a new O chord, we multiply by a factor of $q^{m/2}$ which accounts for all closed chords to its past.\footnote{We have split the factor $\beta q^{-m/2}$ such that $\beta$ only goes with the ``delete'' operator and $q^{m/2}$ goes with the ``add'' operator. This is convenient since one does not know whether an open chord will be right-moving or left-moving. An open $O$ chord becomes a right-mover if it is contracted with $\bq_R$. If it is contracted with $\bq_L$, it will be a left-mover. Thus we only act with $\beta$ or $\beta^\dagger$ when a chord is deleted. This argument also explains why $\beta$ appears in $\qb_R$ whereas $\beta^\dag$ appears in $\qb_L$, see equations \nref{qbrbox} and \nref{qblbox}. }

Now we determine $A,B$. %
Since $Q_R$ acts on the right, if we delete the right-most chord, we will have no intersections. The relevant diagrams are:
\begin{align}
\text{friends:} \qquad  q^{-1/2} \quad \wick[arrows={-,W-<-}]{ \c2{ \color{gray} \qb} \Compactcdots \c2{Q} } \Compactcdots \wick[arrows={-,W-<-}]{\c2{\qb}\c2{\color{lblue} Q} } ,  \qquad 
\text{enemies:} \qquad q^{+1/2} \quad \wick[arrows={-,W->-}]{ \c2{\color{gray} Q} \Compactcdots \c2{\qb} } \Compactcdots \wick[arrows={-,W-<-}]{\c2{\qb}\c2{ \color{lblue} Q} } .  
\end{align}
On the left, we are acting with ${\color{lblue} Q_R}$ on a state of the form $\ket{\cdots \O \cdots \X}$. On the right, we are act on a state of the form $\ket{ \cdots \X \cdots \X}$. In both cases, the rightmost ${\color{lblue} Q}$ represents the operator acting on the state.
We have shaded in gray some supercharge which is not part of the ket vector, but represents a supercharge in any bra vector that can be contracted with the chord state of interest\footnote{E.g., any bra vector that has a non-zero overlap with the specified ket vector must contain at least one such $\bq$ or $Q$.}. %
We get a friendly factor of $q^{-1/2}$ for each O chord and an enemy factor of $q^{1/2}$ for each X chord (except for the X chord we are closing). So this gives $\tilde B \propto q^{(n_\text{X}-1 - n_\text{O})/2}$. 

For the $\tilde A$ term, we need to consider the diagrams:
\begin{align}
& \text{enemies:}  \hspace{1.8cm} -q^{+1/2} &&\wick[arrows={-,W-<-,-,W->-,-}]{ \c4{\color{gray} Q} \Compactcdots \c2{\qb}  \Compactcdots \c4{\qb} \Compactcdots \c2{\color{lblue} Q} } , \quad  && \wick[arrows={-,W-<-,-,W-<-,-}]{ \c4{\color{gray} \qb} \Compactcdots \c2{\qb}  \Compactcdots \c4{Q} \Compactcdots \c2{\color{lblue} Q} }
\end{align}
This gives an intersection and enemy term $(-1)^{n_\text{X} + n_\text{O} -1} q^{(n_\text{X} + n_\text{O} -1)/2} =-(-1)^F q^{(n_\text{X} + n_\text{O} -1)/2}$.

Putting things together, we get
\begin{align}
    A = (-1)^F q^{n_{\tot}/2 }, \quad B= q^{(n_\X - n_\O)/2} \la{abdef}
\end{align}

The conclusion is that we can write 4 supercharges
\def\nt{n_\mathsf{tot}}
\begin{subequations}
\la{supercharges}
\begin{empheq}[box=\widefbox]{align}
Q_\rt &=  \lb (-1)^F q^{\nt/2}  \,\text{del X}_\lt    +    q^{(n_\text{X} - n_\text{O})/2}  \, \text{del X}_\rt \rb \beta^\dagger   + q^{+m/2}\, \text{add O}_\rt \la{qrbox} \\
{\bar Q}_\rt &=  \lb (-1)^F q^{\nt/2}  \, \text{del O}_\lt    +    q^{(n_\text{O} - n_\text{X})/2}  \, \text{del O}_\rt \rb \beta   + q^{-m/2}\, \text{add X}_\rt   \la{qbrbox} \\
(-1)^FQ_\lt  &=  \lb (-1)^F q^{\nt/2}  \, \text{del X}_\rt    +    q^{(n_\text{X} - n_\text{O})/2}  \, \text{del X}_\lt \rb \beta  +q^{-m/2} \, \text{add O}_\lt  \la{qlbox} \\
\bar{Q}_\lt (-1)^F   &=  \lb (-1)^F q^{\nt/2} \,  \text{del O}_\rt    +    q^{(n_\text{O} - n_\text{X})/2}  \, \text{del O}_\lt \rb \beta^\dagger   +  q^{+m/2} \, \text{add X}_\lt \la{qblbox}
\end{empheq}
\end{subequations}
To explain the additional factors of $(-1)^F$ in the left supercharges, let us consider the action of $Q_L$ on a general 2-sided state. Without loss of generality, we can write such a state as $\ket{\psi} = (O_\psi)_R \ket{\Omega}$. %
Now the point is that $Q_L (O_\psi)_R \ket{\Omega} = \pm (O_\psi)_R Q_L \ket{\Omega} = \pm (O_\psi)_R Q_R \ket{\Omega}$, where the $\pm$ sign depends on whether $O_\psi$ is a bosonic or fermionic operator, or equivalently whether $\ket{\psi}$ is a bosonic or fermionic state. %

With these expressions \nref{supercharges} in hand, one can explicitly check that these supercharges satisfy the $\mathcal{N}=4$ SUSY algebra:
\begin{align}
    \{Q_i,Q_j\} &= \{ \bq_i, \bq_j \} = 0, \la{qq0} \\
    \{Q_i, \bq_j\} &= \delta_{ij} H,\la{qqh} \\
    [J_i,Q_j] &= +\delta_{ij} Q_j,\la{u1r1} \\
    [J_i, \bq_j] &= - \delta_{ij} \bq_j \la{u1r2}
\end{align}
As checking these relations by hand is rather tedious, we have included a {\it Mathematica} notebook in the published Supplementary Materials.
Here $i,j$ run over $L,R$. This is the same algebra that is satisfied by the $\mathcal{N}=2$ super-Schwarzian theory in the super-Liouville formalism \cite{Lin:2022rbf}. To verify the last lines, we must find the chord Hilbert space action of the $U(1)_{\mathrm{R}}$ charges, the subject we will now turn to.

\subsection{Action of the \texorpdfstring{$U(1)_{\mathrm{R}}$}{U(1)} charges}
In addition to the action of the supercharges on this Hilbert space, we need to work out the action of the $U(1)_{\mathrm{R}}$ charges. The basic idea is to first rewrite \nref{U1r} as 
\eqn{
J = \lim_{p' \to 1} \frac{1}{\delta \lambda} \lb \mathcal{O} , \bar{\mathcal{O}}\rb , \quad \mathcal{O} = K_{I} \Psi_I, \quad  \delta = p'/p 
} 
Here $\psi_I$ is a product of $p'$ fermions. This might seem like a somewhat cumbersome expression, but the point is that $\bar{\mathcal{O}} \mathcal{O}$ just adds and then immediately deletes a ``matter'' chord.  This ``matter'' chord obeys similar rules except that $q \to q^{\delta}$.
The $p' \to 1$ limit can then be accomplished via a $\delta \to 0$ limit at the end of the derivation\footnote{See \cite{Lin:2022rbf} for a similar derivation of the size operator. The heuristic small $\Delta$ limit was justified explicitly.}.
The composite operator opens and then immediately closes a matter chord. 
Graphically, we can cut open the diagrams on some slice and the insert $({\color{red} \mathcal{O} \bar{\mathcal{O}} } )_R$:
\begin{align}
\label{exampleChordDiagram3}
\begin{tikzpicture}[scale=0.7, baseline={([yshift=0cm]current bounding box.center)}]
	\draw[thick]  (0,0) circle (1.6);
\begin{scope}[thick, decoration={markings, mark=at position 0.2 with {\arrow{>}}, mark=at position 0.9 with {\arrow{>}} }]	
        \draw[postaction=decorate]  (0,1.6) node[above] {$Q$} -- (-1.13,-1.13)  node[below left] {$\bar Q$};
	\draw[postaction=decorate] (1.55025987,-0.39584) node[right] {$Q$} --(-1.05,1.2) node[above left] {$\bar Q$};
	\draw[postaction=decorate] (-1.58,-0.2) node[left] {$Q$} -- (1.13,-1.13) node[right] {$\bar Q$};
	\draw[postaction=decorate] (0,-1.6) node[below] {$Q$} -- (0.93069294,1.30146481) node[above right] {$\bar Q$};
 \end{scope}
	\draw[fill,black] (-1.58,-0.2) circle (0.1);
	\draw[fill,black] (-1.05,1.2) circle (0.1);
	\draw[fill,black] (1.55025987,-0.39584) circle (0.1);
	\draw[fill,black] (0.93069294,1.30146481) circle (0.1);
	\draw[fill,black] (-1.13,-1.13) circle (0.1);
	\draw[fill,black] (1.13,-1.13) circle (0.1);
	\draw[fill,black] (1.13,-1.13) circle (0.1);
	\draw[fill] (0,-1.6) circle (0.1);
	\draw[fill] (0,1.6) circle (0.1);
        \draw[dashed] (-1.6,0) -- (1.6,0);
        \draw[fill,gray] (-1.6,0) circle (0.1);
	\draw[fill,gray] (1.6,0) circle (0.1);
\end{tikzpicture} \qquad \RA \qquad 
\begin{tikzpicture}[scale=0.7, baseline={([yshift=0cm]current bounding box.center)}]
	\draw[thick]  (0,0) circle (1.6);
\begin{scope}[thick, decoration={markings, mark=at position 0.5 with {\arrow{>}}}]	
        \draw[postaction=decorate]  (0,1.6) node[above] {$Q$} -- (-1.13,-1.13)  node[below left] {$\bar Q$};
	\draw[postaction=decorate] (1.55025987,-0.39584) node[right] {$Q$} --(-1.05,1.2) node[above left] {$\bar Q$};
	\draw[postaction=decorate] (-1.58,-0.2) node[left] {$Q$} -- (1.13,-1.13) node[right] {$\bar Q$};
	\draw[postaction=decorate] (0,-1.6) node[below] {$Q$} -- (0.93069294,1.30146481) node[above right] {$\bar Q$};
        \draw[red,postaction=decorate]  (1.4,0.8) node[above right] {$\mathcal{O}$} --   (1.55,0.05) node[above right] {$\bar{\mathcal{O}}$}  ;
 \end{scope}
	\draw[fill,black] (-1.58,-0.2) circle (0.1);
	\draw[fill,black] (-1.05,1.2) circle (0.1);
	\draw[fill,black] (1.55025987,-0.39584) circle (0.1);
	\draw[fill,black] (0.93069294,1.30146481) circle (0.1);
	\draw[fill,black] (-1.13,-1.13) circle (0.1);
	\draw[fill,black] (1.13,-1.13) circle (0.1);
	\draw[fill,black] (1.13,-1.13) circle (0.1);
	\draw[fill] (0,-1.6) circle (0.1);
	\draw[fill] (0,1.6) circle (0.1);
        \draw[fill,red] (1.4,0.8) circle (0.1);
        \draw[fill,red] (1.55,0.05) circle (0.1);
\end{tikzpicture}
\end{align}
Similar reasoning\footnote{For example, to derive the action of $\Psi_{I,R} \bar{\Psi}_{I,R}$ we consider the action of the ``add'' term in \nref{qbrbox} followed by the ``del'' term in \nref{qrbox} and take $q \to q^\delta$. } to \nref{supercharges} gives the operator expressions %
\begin{align}
    (\Psi_I \bar{\Psi}_I)_\rt &=  \frac12  \beta^{-\delta} e^{ - \hf \lambda \delta (n_\text{X}-n_\text{O} -m)}  , \quad
     (\Psib_I \Psi_I)_\rt  =  \frac12 \beta^{\delta} e^{ - \hf \lambda\delta  (n_\text{O}-n_\text{X}+m)}, \la{ppr}\\
     (\Psi_I \bar{\Psi}_I)_\lt &=\frac12  \beta^{\delta} e^{ - \hf \lambda \delta (n_\text{X}-n_\text{O} + m)}  , \quad
     \; \; (\Psib_I \Psi_I)_\lt  =\frac12 \beta^{-\delta} e^{ - \hf \lambda \delta (n_\text{O}-n_\text{X} - m)} \la{ppl} 
\end{align}
In the above expressions, we are imagining acting with $\mathcal{O}, \bar{\mathcal{O}}$ on a state with no other insertions of $\mathcal{O}, \bar{\mathcal{O}}$. We have included a factor of $\hf$ to match the normalization of the Dirac fermions \nref{dirac}. Then performing the disorder average contracts the factors of $K$ to give the LHS of equations \nref{ppr} and \nref{ppl}.
Then taking the limit $\delta \to 0$, we can extract
\begin{align}
    [\Psib_I, \Psi_I]_\rt \sim \delta \lb \frac{\lambda}{2} (n_\text{X}-n_\text{O} - m)  + \pd_m \rb \\
    [\Psib_I, \Psi_I]_\lt \sim \delta \lb \frac{\lambda}{2} (n_\text{X}-n_\text{O} + m)  - \pd_m \rb 
\end{align}
Here we view $\beta^\delta$ as acting on wavefunctions via $e^{\delta \pd_m} \psi(m) = \psi(m+\delta)$. %
The conclusion is that the $U(1)_{\mathrm{R}}$ charges are %
\begin{align}
    J_\rt &=  \frac12 (n_\text{O} - n_\text{X} + m) - \frac{1}{\lambda} \pd_m  ,\\
    J_\lt &=  \frac12 (n_\text{O} - n_\text{X} -m) + \frac{1}{\lambda} \pd_m .
\end{align}
One can then check that the supercharges have the appropriate $U(1)_\mathrm{R}$ charges, \nref{u1r1} and \nref{u1r2}.

\subsection{Diagonalizing the \texorpdfstring{$U(1)_\mathrm{R}$}{U(1)} charges\label{diagonalCharge}}
Since $J_\lt, J_\rt$ commute with the Hamiltonian, we may simultaneously diagonalize $J_L \pm J_\rt$ and the Hamiltonian $H$.
The combination $J_L + J_R$ is trivial to diagonalize: one can consider eigenstates with fixed $n_\text{X} - n_\text{O}$.
Diagonalizing $J_L - J_R$ is slightly less trivial, since
\begin{align}
    J_\lt - J_\rt &= -m +2 \lambda\inv  \pd_m  %
\end{align}
involves both $m$ and its conjugate momentum. Nevertheless, diagonalizing such an operator is familiar from the usual discussion of coherent states; we get:
\begin{align}
\ket{j} = \sum_{m=-\infty}^\infty q^{ -\tfrac14  m^2 - \hf jm}  \ket{m } , \quad (J_L - J_R) \ket{j} = j \ket{j}.
\end{align}

So working in this new eigenbasis, we have
\begin{align}
    \sum_{m=-\infty}^\infty q^{m/2} \psi_j(m) \ket{m} &= \ket{j-1}, 
    \quad 
    \beta^\dag \ket{j} %
    = q^{j/2 - 1/4} \ket{j-1} ,
    \\
    \sum_{m=-\infty}^\infty q^{-m/2} \psi_j(m) \ket{m} &= \ket{j+1}, 
    \quad 
    \beta \ket{j} %
    = q^{-(j+1)/2 + 1/4} \ket{j+1} .
\end{align}

We may write this as $\beta^\dagger=\tilde\beta q^{j/2 - 1/4}$ and $\beta= {\tilde \beta}^\dagger q^{-j/2 - 1/4}$ 
where $\tilde\beta^\dagger \ket{j} = \ket{j+1}, \tilde\beta \ket{j} = \ket{j-1}$.
So we may also rewrite the supercharges as
\begin{subequations}
\begin{empheq}[box=\widefbox]{align}
Q_\rt &=  \tilde{\beta} \;\lb (-1)^F q^{\nt/2} q^{j/2-1/4}  \,\text{del X}_\lt    +    q^{(n_\text{X} - n_\text{O})/2} q^{j/2-1/4} \, \text{del X}_\rt    + \, \text{add O}_\rt \rb \label{jqrt},\\
{\bar Q}_\rt &=  \tilde{\beta}^\dag \! \lb (-1)^F q^{\nt/2} q^{-j/2-1/4} \, \text{del O}_\lt    +    q^{(n_\text{O} - n_\text{X})/2} q^{-j/2-1/4} \, \text{del O}_\rt    + \, \text{add X}_\rt \rb \label{jqrtb},\\
(-1)^FQ_\lt  &=  \tilde\beta^\dag\! \lb  (-1)^F q^{\nt/2}  q^{-j/2-1/4}\, \text{del X}_\rt    +    q^{(n_\text{X} - n_\text{O})/2}  q^{-j/2-1/4}\, \text{del X}_\lt   + \text{add O}_\lt  \rb  ,\\
\bar{Q}_\lt (-1)^F   &=  \tilde\beta \; \lb (-1)^F q^{\nt/2} q^{j/2-1/4}  \,  \text{del O}_\rt    +    q^{(n_\text{O} - n_\text{X})/2}  q^{j/2-1/4}  \, \text{del O}_\lt    +   \text{add X}_\lt \rb \la{jqltb} 
.
\end{empheq}
\end{subequations}
And the left and right $U(1)_\mathrm{R}$ charges:
\eqn{J_L &= \hf (n_\text{O} -n_\text{X} - j), \qquad J_R = \hf (n_\text{O} - n_\text{X} +j). }
Here $\tilde\beta^\dagger \ket{j} = \ket{j+1}$.
Let us write a general state 
\eqn{
 \ket{\psi} =  \sum_{n=0}^\infty \sum_{j=-\infty}^\infty \psi_{\O\O,j} (n) \ket{n,\O\O,j} + \psi_{\O\X,j} (n) \ket{n,\O\X, j} \\
 \hspace{3.3cm} + \psi_{\X\O,j} (n)  \ket{n,\X\O,j} +  \psi_{\X\X,j} (n)\ket{n, \X\X,j}
}
or equivalently as a 4-component wavefunction:
\begin{align}
    \ket{\psi} = \begin{bmatrix}{}
          \psi_{\O\O,j} \lp {n}\rp  \\
          \psi_{\O\X,j} \lp {n}\rp \\
          \psi_{\X\O,j} \lp {n} \rp \\
          \psi_{\X\X,j} \lp {n} \rp \\
    \end{bmatrix} .
\end{align}
In this basis, $Q_R$ acts as 
\begin{subequations}
\begin{align}
  Q_\rt \ket{\psi,j} &=  
\left[
\begin{array}{c}
 q^{\frac{j}{2}-\frac{1}{4}} \left(\psi _{\text{OX},j+1}(n+1)-q^{n+1} \psi _{\text{XO},j+1}(n+1)\right)+\psi _{\text{OX},j+1}(n) \\
 q^{\frac{j}{2}+n+\frac{1}{4}} \psi _{\text{XX},j+1}(n) \\
 q^{\frac{j}{2}+\frac{1}{4}} \psi _{\text{XX},j+1}(n)+\psi _{\text{XX},j+1}(n-1) \\
 0 \\
\end{array}
\right]
\label{qrtmatrix}\\
  \bq_\rt \ket{\psi,j}  &=  
\left[
\begin{array}{c}
 0 \\
 q^{\frac{1}{4}-\frac{j}{2}} \psi _{\text{OO},j-1}(n)+\psi _{\text{OO},j-1}(n-1) \\
 q^{-\frac{j}{2}+n+\frac{1}{4}} \psi _{\text{OO},j-1}(n) \\
 q^{-\frac{j}{2}-\frac{1}{4}} \left(\psi _{\text{XO},j-1}(n+1)-q^{n+1} \psi _{\text{OX},j-1}(n+1)\right)+\psi _{\text{XO},j-1}(n) \\
\end{array}
\right] \la{qrtmatrix2}
\end{align}
\end{subequations}

Let us make a few comments about the thermofield double. 
Note that the symmetric combination $J_\lt + J_\rt =n_\text{X} - n_\text{O}$. For the thermofield double/Hartle-Hawking states, $J_\lt = -J_\rt$ so this combination should annihilate the state. For general empty wormhole states (wormholes with no operator insertions besides $Q, \bar{Q}$), we see that $J_\lt + J_\rt$ has  eigenvalues $\{-1,0,0,+1\}$. The 0 is listed twice because the states $\ket{n, \X\O}$ and $\ket{n,\O\X}$ are both eigenstates with 0 total charge. The fact that there are 4 states is related to the statement that for a boundary system with $\mathcal{N} =2$ SUSY, the thermofield double comes in a multiplet with 4 states \cite{LongPaper}.

So if we set $J_\lt + J_\rt = 0$, we set $n_\text{X} = n_\text{O}$ which implies $\psi_{\X\X} = \psi_{\O\O} = 0$. Furthermore, the one-sided charge on the right $j_R = -j/2$. To compare with BBNR \cite{superBerkooz}, one should take $s =  j_R = -\hf j$ and to compare with LMRS \cite{LongPaper}, one should take $j = 2j_\text{LMRS} $.

\section{Constructing the supersymmetric wormhole \la{susyHartleHawking} }

In this section we use the bulk supercharges constructed above to find the 0-energy ground state of the chord Hamiltonian. This is the analog of the Hartle-Hawking wavefunction in $\mathcal{N}=2$ JT gravity, which is a bound state eigenfunction of the super-Liouville quantum mechanics \cite{Lin:2022rzw,LongPaper}.

\subsection{Finding the Hartle-Hawking ground state}

It will be convenient to work with rescaled wavefunctions $\psi_{\X\O,j}(n) = q^{-n/4} \alpha_n$ and $\psi_{\O\X} = q^{-n/4} \beta_n$.
(See \nref{alphaDef} and \nref{betaDef} in Appendix \ref{innerProductM} for an explanation of the $q^{-n/4}$ factor, which we introduced purely for the reader's convenience, e.g., to enable easy comparison with BBNR \cite{superBerkooz}.)
We consider the following ansatz for the wavefunction for the BPS wormhole: %
\be
\ket{\Psi, j}=\sum_{n=0}^\infty  q^{-n/4} \lp \alpha_n\ket{n, \X\O, j}+ \beta_n\ket{n, \O\X, j} \rp .\la{ansatzBPS}
\ee
Setting $Q_\rt \ket{\psi,j} =\qb_\rt \ket{\psi,j} = 0$ and using \nref{qrtmatrix} and \nref{qrtmatrix2} gives
\begin{align} 
-q^{n-1} \alpha_{n} + q^{-1} \beta_{n}    + q^{j_R}  \beta _{n-1}  &= 0\label{gsBerkooz1} 
,
\\
q^{-1}  \alpha_{n} + q^{-j_R}\alpha_{n-1} - q^{n-1}  \beta_{n} &=0\label{gsBerkooz2}
.
\end{align}

We can decouple the two wavefunctions:
\begin{align}
(\q^{-3}-\q^{2n-1})\alpha_{n+1} + (\q^{-2-j_R}+\q^{j_R-1}) \alpha_n + \alpha_{n-1} &= 0 ,
\label{alphaeq}\\
(\q^{-3}-\q^{2n-1})\beta_{n+1} + (\q^{-2+j_R}+\q^{-j_R-1}) \beta_n + \beta_{n-1} &= 0 .\la{betaeq}
\end{align}
With the initial condition\footnote{Any choice of non-zero $\alpha_0$ and $\beta_0$ will lead to the same physical predictions. A different choice of initial condition only rescales the wavefunction, which will be taken care of later by normalizing the wavefunction using the inner product.} $\alpha_0 = \beta_0 =1$, these recursion relations can then be solved using q-Hermite polynomials:
\begin{align}
\alpha_n & =\frac{\q^{\frac{3n}{2}}}{(\q^2;\q^2)_n}H_n\left[-\cosh\left(\lambda(j_R+\hf)\right) \big|\q^2\right], \\
\beta_n &=\frac{\q^{\frac{3n}{2}}}{(\q^2;\q^2)_n}H_n\left[-\cosh\left(\lambda(j_R-\hf)\right) \big|\q^2\right] . 
\end{align}
Notice that the equations are invariant under $\alpha\leftrightarrow\beta$ and $j_R\leftrightarrow-j_R$.

\begin{figure}
    \centering
    \includegraphics[width=0.7\columnwidth]{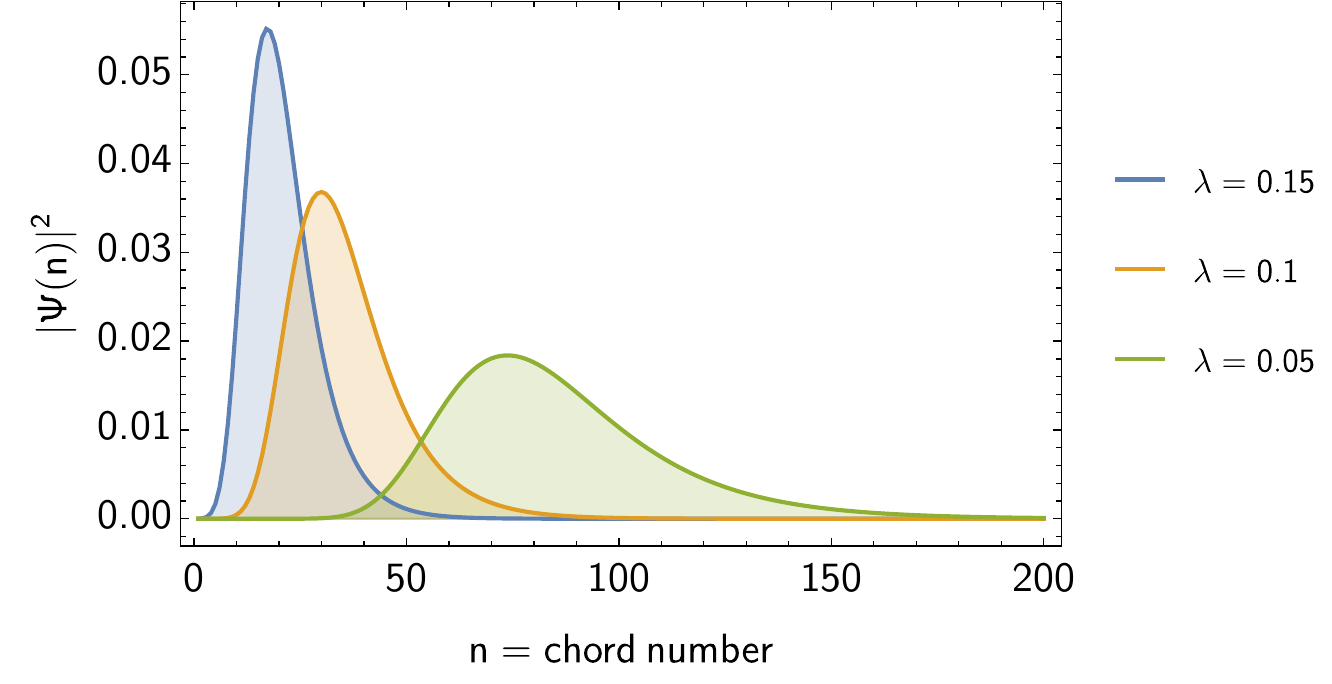}
    \caption{Wavefunction of the $j=0$ ground state. We plot the probability that the wormhole has some fixed chord number. As $\lambda$ decreases, the wavefunction recedes to larger values of $n$. The probability that the wormhole has zero length $|\Psi(0)|^2$ gives the {\it fraction} of states in the Hilbert space that are supersymmetric. In the small $\lambda$ regime this agrees with the super-Schwarzian prediction.}
    \label{fig:probChord}
\end{figure}

We have computed the wavefunctions of the ground state, but to complete the discussion, we should compute the probability distribution for the length of the wormhole.
To do so, we need the inner products between various states, which we derive in 
Appendix \ref{innerProductM}, and display here:
\begin{align}
\braket{n, \O\X, j}&=  \braket{n, \X\O, j}=\q^{n/2}\left(\q^2 ; \q^2\right)_{n-1} \la{inner1} \\
\braket{n, \O\X, j}{n, \X\O, j} &=-q^{3n/2} \left(\q^2 ; \q^2\right)_{n-1} \la{inner2}
\end{align}
Using this, we can compute the norm of the ground state is given by
\begin{align}
\la{eq:norm}
\bra{\Psi, j}\ket{\Psi, j} &= 
\sum_n (\q^2; \q^2)_{n-1}  \lb \q^{-n} \lp \alpha_n \alpha_n + \beta_n \beta_n \rp - 2 \alpha_n \beta_n \rb 
\\
&= \frac{1}{(\q^{1 \pm 2j_R};\q^2)_\infty (\q^2;\q^2)_\infty}. \la{eq:norm2}
\end{align}
The last line is a $q$-identity that we prove in Appendix \ref{app:norm}. If the $\pm$ sign only appears on one side of an equation, we are instructed to take a product, e.g. $f(x \pm y) = f(x+y) f(x-y)$. Furthermore, the term that appears in the sum \nref{eq:norm} divided by the norm \nref{eq:norm2} can be interpreted the probability $P_n$ that the wormhole has chord number $n$. We plot this probability in Figure \ref{fig:probChord} for various values of $\lambda$. At small $\lambda$, the probability distribution is peaked at large chord number, which means that the BPS wormhole is typically quite long.

In Appendix \ref{schMatch}, we consider the $\lambda \to 0$ (``triple scaling'') limit of these expressions and compare them with the $\caln = 2$ super-Schwarzian wavefunctions computed in \cite{LongPaper}. We find precise agreement.

\subsection{A bulk computation of the ground state degeneracy\la{sec:bulkComp}}
As a first application of this wavefunction, we will compute the fraction $D$ of microstates that are zero energy\footnote{Do not confuse the various notions of ground states. The ground state of the ``bulk'' or ``chord'' Hamiltonian is a 2-sided concept. It is the bound state wavefunction we are discussing in \nref{susyHartleHawking} that represents the length of the wormhole. There is a unique ground state for each value of $|j_R| < 1/2$. On the other hand, there are exponentially many zero-energy ground states of a single copy of the microscopic theory. The bulk description of such microstates, if it exists, would be one-sided geometries, not the wormhole that we have been discussing. These 1-sided microstates are what is being counted by the fraction $D(j_R)$.} (extremal black hole states):%
\be
D(j_R)\doteq \frac{ \dim \mathcal{H}_{\text{extremal}, \, j_R} }{ \dim \mathcal{H}_{j_R} }=\frac{Z(j_R, \infty)}{Z(j_R, 0) }
.
\ee
To do so, we will use \nref{superBowlAd} as advertised in the introduction. The derivation of \nref{superBowlAd} is extremely simple using the ingredients we have already encountered. Taking $\beta \to \infty$ of \nref{micro}, we find the chord Hilbert space expression %
\be
Z(j_R, \infty)=\frac{|\braket{\Omega,j}{\Psi,j}|^2}{\braket{\Psi,j} } .
\ee
On the other hand, taking $\beta \to 0$ of \nref{micro}, %
\be
Z(j_R, 0)=\braket{\Omega,j} .
\ee
Thus the fraction is given by
\be
D(j_R) = \frac{|\braket{\Omega,j}{\Psi,j}|^2}{\braket{\Omega,j} \braket{\Psi,j} } . \la{superBowl2}
\ee
As promised in \nref{superBowlAd}, the RHS is precisely the probability that the wormhole has zero length, since $\Omega$ is the state with the smallest number of chords $n=0$. %

The simplicity of the derivation also belies its generality, as it made very little reference to the details of the chord theory. Just to emphasize this point, let us present essentially the same derivation again using pictures and words associated to Euclidean quantum gravity: %
\begin{align}
|\braket{\text{susy wormhole} \, }{\ell = 0}|^2 \; = \;\lim_{\beta\to \infty} \bra{\ell = 0} e^{-\beta H} \ket{\ell = 0} \; = \; \vcenter{ \hbox{\includegraphics[scale=0.4]{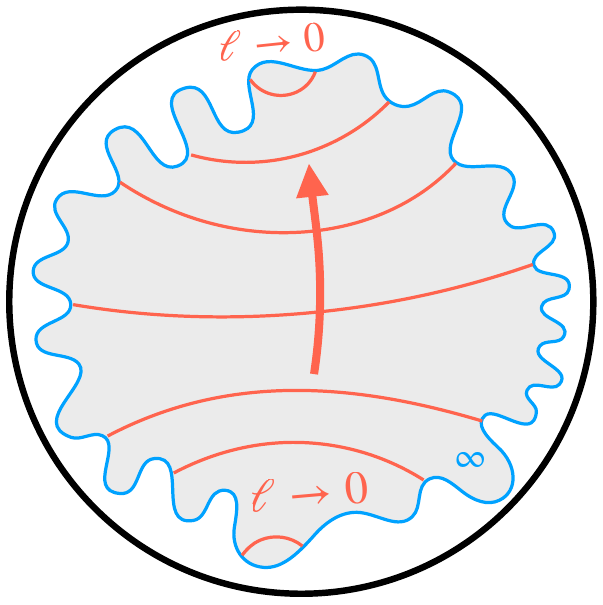}}} \la{intro2}
\end{align}
Start with the middle expression. To understand the equality with the RHS of \nref{intro2}, note that the middle expression can be viewed as an amplitude from a short ``$\ell=0$'' wormhole
in the distant Euclidean past to the same short wormhole in the Euclidean future. This is just a way of slicing up the disk partition function, which as $\beta \to \infty$ gives the zero temperature partition function $Z_{\beta= \infty} = \dim \mathcal{H}_{\text{extremal}}$. On the other hand, if we evolve for very large $\beta$, we should project to the zero energy state of the bulk Hamiltonian. This is the Hartle-Hawking wavefunction of a supersymmetric wormhole (e.g. a wormhole joining two extremal black holes), which in the JT gravity approximation is described by a bound state wavefunction in the length mode \cite{Lin:2022rzw, LongPaper}. This gives the LHS of \nref{intro2}. Now dividing both sides by $\braket{\ell = 0}$ and interpreting $\braket{\ell = 0}$ as the infinite temperature partition function $Z = \dim \mathcal{H}$ gives \nref{superBowlAd}.

The only subtle difference between the ``gravity'' derivation presented above and the superchord derivation is that the $\ell = 0$ state has a precise meaning in the superchord theory, whereas in a theory like JT gravity the $\ell = 0$ does not have a sharp meaning. To be more precise, in JT gravity, we should always work with the renoramlized length; the statement that $\ell \to 0$ is really the statement $\tilde \ell \to -\infty$. In this limit, we are sensitive to what the precise UV completion of JT gravity is; see \cite{Lin:2022rbf} for more discussion\footnote{This is closely related to the fact that the algebra of one-sided observables in JT gravity is Type II$_\infty$ \cite{Penington:2023dql,Kolchmeyer:2023gwa} whereas in DSSYK it is Type II$_1$. Hence the maximally entangled state (the state of maximum entropy) is a well-defined concept in DSSYK but it is not well-defined in JT.}.

\begin{figure}[H]
    \centering
    \includegraphics[width=0.7\columnwidth]{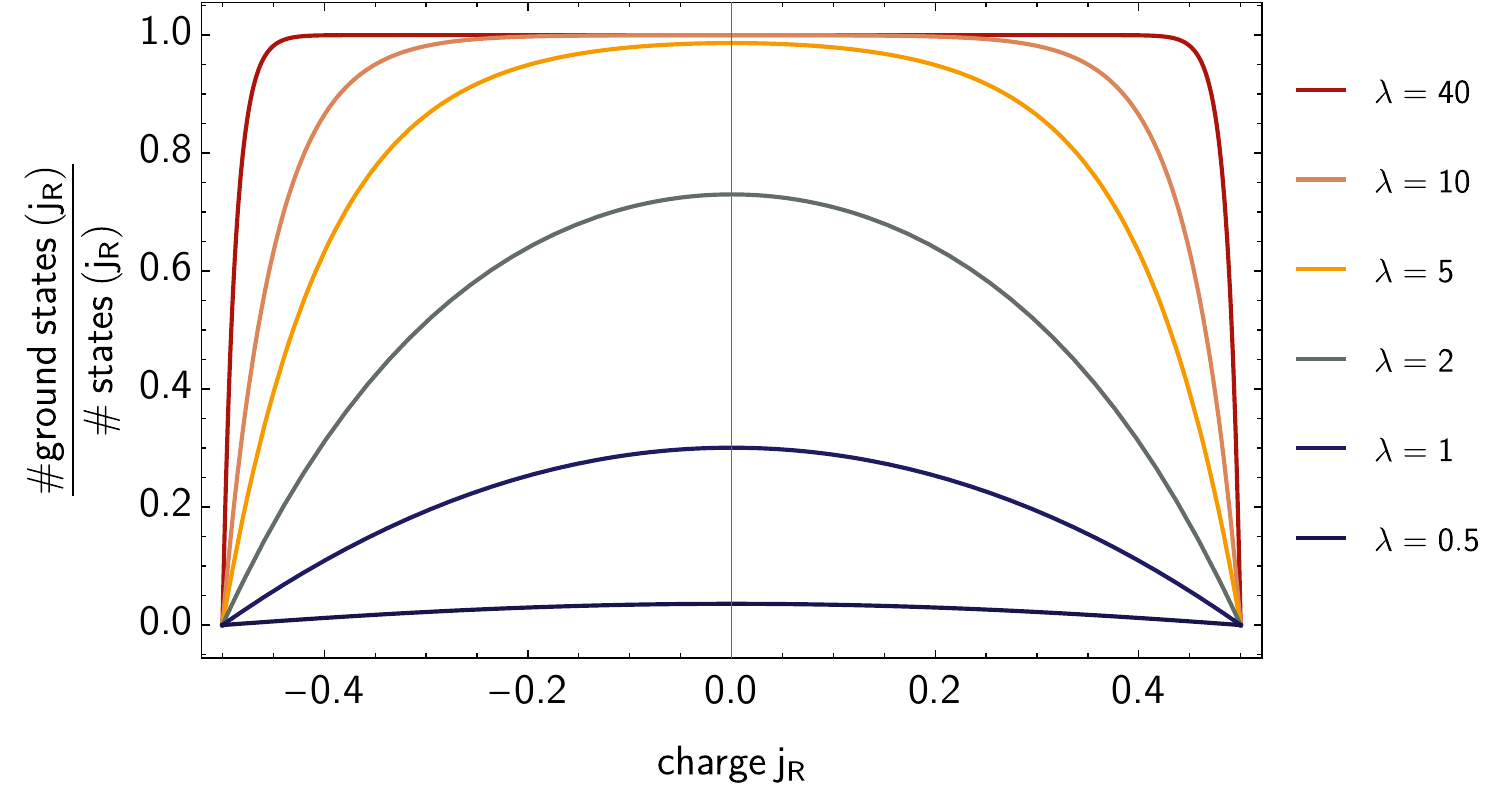}
    \caption{Fraction of ground states as a function of charge $j_R$ for different values of $\lambda$. For small $\lambda$, the number of ground states is a tiny fraction of the total number of states of charge $j_R$.}
    \label{fig:gs_fraction}
\end{figure}

Now let's return to explicitly evaluating $D(j_R)$ using \nref{superBowl2}.
Using \nref{eq:norm} and the expressions for the inner products \nref{inner1}-\nref{inner2} where the $\Omega$ state corresponds to $n=0$, 
we arrive at
\begin{align}
D(j_R) &=(\q^{2j_R+1};\q^2)_\infty(\q^{-2j_R+1};\q^2)_\infty(\q^2;\q^2)_\infty \\
&=\vartheta_4(\i\lambda j_R,\q)\\
&=\sqrt{\frac{\pi}{\lambda}}\q^{-j_R^2}\vartheta_2(\pi j_R,e^{-\frac{\pi^2}{\lambda}}) \la{bulkD}
.
\end{align}
We would like to compare this to the microscopic expectations, coming from the index. To this end, let us define another fraction, where we normalize by the total number of states in the entire SYK Hilbert space:
\be
\hat{D}(j_R)=\frac{ \dim \mathcal{H}_{\text{extremal}, \, j_R}}{\dim \mathcal{H}  } %
.
\ee
As reviewed in Appendix \ref{app:index}, the finite $N$ microscopic refined index computations give
\be
p \hat{D}(j_R)= 2  \cos \left({\pi j_R}\right)\left[ \cos \frac{\pi}{2 p }\right]^N  + 2 \cos (3 \pi j_R) \lb \cos \frac{3\pi}{2 p} \rb^N + 2\cos (5 \pi j_R) \lb \cos \frac{5\pi}{2 p} \rb^N + \cdots \la{microCount}
\ee
To convert $\hat{D} \to D$, we note that $\dim \mathcal{H} = 2^N$, while the total number of states of charge $j_R$ is 
\be
\frac{\dim \mathcal{H}_{j_R} }{\dim \mathcal{H}}= 2^{-N} \binom{ N}{ N/2 + j_R p } \approx \frac{1}{p} \sqrt{\frac{\lambda}{\pi}}e^{-\lambda j_R^2}. \la{chargefactor}
\ee
In the last step, we took the double scaling limit by approximating the binomial distribution with a Gaussian. Taking the double scaling limit of \nref{microCount} and then multiplying by \nref{chargefactor} gives
\be
D(j_R)= \sqrt{\frac{\pi}{\lambda} }q^{-j_R^2}  \left[  2 \cos (\pi j_R)  e^{- \pi^2/(4\lambda)} + 2 \cos (3\pi j_R)e^{- 9\pi^2/(4\lambda)}  +  2 \cos (5 \pi j_R ) e^{-25 \pi^2/(4\lambda) } + \cdots \right]
.
\ee
This agrees precisely with the expression we derived from the bulk in \nref{bulkD}. We plot this fraction in Figure \ref{fig:gs_fraction} for various values of $\lambda$ and various charges $j_R$.

Note that the leading $j_R$-dependence given by $\cos \pi j_R$ is predicted by the super-Schwarzian \cite{Stanford:2017thb}. To get the other terms, we needed to go beyond the Schwarzian theory and use a better bulk description. Here the better description was the super chord theory. 

We can also understand the fraction of ground states in the extreme double scaling limit $\lambda \gg 1$. Recall that in an $\nn = 2$ theory
\eqn{\# \text{ ground states} =  \mathop{\mathrm{ker}} Q / \mathop{\mathrm{im}} Q .}
According to \nref{chargefactor}, the dimension of the Hilbert space is proportional to $e^{- \lambda j_R^2}$. This Gaussian has a width of $\sim 1/\sqrt{\lambda}$ which is very narrow in the large $\lambda$ limit. Acting on a microstate, $Q$ increments $j_R \to j_R + 1$.
Suppose we consider a state with $|j_R| < 1/2$. For $j_R > - 1/2$, the number of states with charge $j_R + 1$ is much less than the number of states with charge $j_R$ in the $\lambda \gg 1$ limit. Hence $Q$ must have a large kernel.
On the other hand, consider states of charge $j_R$ that can be written as $\ket{\chi,j_R} = Q \ket{\chi', j_R - 1}$. For $j_R < 1/2$, the number of states with charge $j_R - 1$ is quite small compared to the total number of states with charge $j_R$. Hence we may neglect the number of states that are $Q$-exact. Thus we conclude that \\
\eqn{ \text{for $|j_R| < 1/2$, nearly all states are BPS in the $\lambda \gg 1$ limit},}
which is indeed what we see in Figure \ref{fig:gs_fraction}. This argument also predicts that the convergence to $0$ in Figure \ref{fig:gs_fraction} should be slower near\footnote{The number of states of charge $j=-1/2 + \epsilon$ is close to $j = 1/2+\epsilon$ unless $\epsilon \lambda \gg 1$.} the extreme points $j_R \to \pm 1/2$, which is indeed what we observe.

\section{Zero temperature 2-pt function and the length of the wormhole \label{sec:2pt}} 
\def\Gammaq{\Gamma_{q^2}}
In this section, we consider the zero temperature 2-pt function. More precisely, we consider correlators such as $\tr \lp \Pi_{0,j_R'}  \vb \, \Pi_{0,j_R} \V \rp $ or $\tr \lp \Pi_{0,j_R} \W \, \Pi_{0,j_R} \W \rp $ where $\Pi_{0,j_R}$ is the projector onto the ground state sector of charge $j_R$. Such correlators also govern the statistics of the 1-pt function in a zero energy microstate \cite{Lin:2022rzw, LongPaper}. One can use the 2-sided Hilbert space to compute such correlators. In particular, we can view a correlator like $\tr \lp \Pi_{0,j_R} \W \, \Pi_{0,j_R} \W \rp = \bra{\Psi,j} \W_L \W_R \ket{\Psi,j}$ and then view this as the action of a ``horizontal'' matter chord on the Hartle-Hawking wavefunction that we have already computed.

\subsection{Horizontal matter chords}
Now we derive the action of a ``horizontal'' matter chord.
We consider the matrix elements of $\vb_L \V_R$ between arbitrary 2-sided states $\ket{\Psi,j}$ and $\bra{\Phi,j'}$. This will be given by a sum over chord diagrams:
\begin{align}
\label{exampleChordDiagramMatter}
\bra{\Phi,j'} \vb_L \V_R \ket{\Psi, j} \quad \supset \quad  \begin{tikzpicture}[scale=0.7, baseline={([yshift=0cm]current bounding box.center)}]
	\draw[thick]  (0,0) circle (1.6);
\begin{scope}[thick, decoration={markings, mark=at position 0.5 with {\arrow{>}} }]	
        \draw[postaction=decorate] (-1.4,-0.75) -- (0.5,-1.5);
        \draw[postaction=decorate]  (0,1.6) -- (-1.13,-1.13)  ;
        \draw[postaction=decorate]  (0,1.6) -- (-1.13,-1.13)  ;
	\draw[postaction=decorate] (1.55025987,-0.39584) --(-1.05,1.2) ;
	\draw[postaction=decorate] (-1.58,-0.2) -- (1.13,-1.13) ;
	\draw[postaction=decorate] (0,-1.6) -- (0.93069294,1.30146481) ;
        \draw[postaction=decorate, red] (1.6,0) -- (-1.6,0);
 \end{scope}
	\draw[fill,black] (-1.58,-0.2) circle (0.1);
	\draw[fill,black] (-1.05,1.2) circle (0.1);
	\draw[fill,black] (1.55025987,-0.39584) circle (0.1);
	\draw[fill,black] (0.93069294,1.30146481) circle (0.1);
	\draw[fill,black] (-1.13,-1.13) circle (0.1);
	\draw[fill,black] (1.13,-1.13) circle (0.1);
	\draw[fill,black] (1.13,-1.13) circle (0.1);
	\draw[fill] (0,-1.6) circle (0.1);
	\draw[fill] (0,1.6) circle (0.1);
        \draw[fill,red] (-1.6,0) circle (0.1);
	\draw[fill,red] (1.6,0) circle (0.1);
        \draw[fill] (-1.4,-0.75) circle (0.1);
        \draw[fill] (0.5,-1.5) circle (0.1);
\end{tikzpicture}.
\end{align}
We can leverage our experience from working with the supercharges to quickly derive the action of $\vb_L \V_R$ 
in the basis specified in Section \ref{diagonalCharge}. 
We consider $(\bq_L)_{\text{del}} (Q_R)_{\text{add}} $ and then replace $q \to q^{\Delta_\O }$. Here we only consider the term in \nref{jqltb} that contains $(\text{del O}_R)$, since we want the term that deletes the chord that $(Q_R)_{\text{add}}$ created. This gives
\begin{align}
   \vb_L \V_R  &\propto - {\tilde\beta}^{2\Delta_\O } q^{\hf \Delta_\O (j+n_\O + n_\X)} \la{vbv} ,\\
   \V_L \vb_R  &\propto  ({\tilde{\beta} ^\dagger})^{2\Delta_\O} q^{\hf \Delta_\O (-j+n_\O + n_\X)} \la{vbv2}  .
\end{align}
These expressions hold on the subspace of states which are empty wormholes. We have assumed that $\V$ is a fermionic operator. 
(If $\V$ is bosonic, we would not have the minus sign in \nref{vbv}.)

We can also consider a more general operator $\W$ that is not BPS. 
We will focus on the case where $\W$ is neutral; it is easy to extend our results to the general charge with these two cases in hand. 
Heuristically, a neutral operator is the product of a chiral and anti-chiral operator $\W \sim \vb_1 \V_2$ where $\vb_1$ and $\V_2$ are independent BPS operators. 
Then we can write the 2-pt function as $(\vb_L \V_R)(\V_L \vb_R) $. Then using \nref{vbv} and \nref{vbv2} yields %
\begin{align}
    \W_L \W_R \propto  q^{\Delta (n_\O + n_\X ) }
    .
\end{align}
One can also justify this expression directly from the chord rules, without appealing to the heuristic $\W \sim \vb_1 \V_2$.
Note that there is a proportionality constant that is depends on our microscopic choice of normalization for $\V$ and $\W$. We choose a convention where we set the proportionality constant to unity. %

\subsection{2-pt functions and the length of the wormhole}

\subsubsection*{2-pt function of neutral operators}
Following the above discussion, to evaluate a thermal correlation function with two neutral matter operator insertions $\mathcal{O}_\Delta$ in the $U(1)_{\mathrm{R}}$ charge sector $j_R$, the chord rules instruct us to insert an additional factor of $q^{2\Delta n}$ in the equation \nref{eq:norm}. We are therefore interested in evaluating the expression of the form 
\be
\ev{\W_L \W_R}{\Psi, j} = \bra{\Psi,j} q^{\Delta ({n}_{\X} + {n}_{\O})} \ket{\Psi,j} ,
\ee
which after plugging in the ground state wavefunction and using the inner product \nref{inner1} takes the form
\be 
\bra{\Psi,j} q^{2\Delta n} \ket{\Psi,j} 
= 
\sum_{n=0}^\infty {\q^{2\Delta n}}    {(q^2 ; q^2)_{n-1}} (\q^{-n}(\alpha_n^2 + \beta_n^2) - 2 \alpha_n \beta_n)
.
\ee
The above expression is similar to the expressions for the two-point functions in JT gravity using the Liouville formalism \cite{Bagrets:2016cdf, Harlow:2018tqv}, by gluing together two Hartle-Hawking wavefunctions along a common geodesic $\ell$, with an additional insertion of a factor $e^{-\Delta \ell}$ \cite{Yang:2018gdb}. The above expression is precisely a generalization of that computation away from the JT gravity limit, with the bulk ground state $\ket{\Psi,j}$ being the generalized ground sector Hartle-Hawking state, and the chord number $n$ being related to geodesic length $\ell$ via $\ell = 2\lambda n$.

\begin{figure}
\centering
\includegraphics[width=0.7\columnwidth]{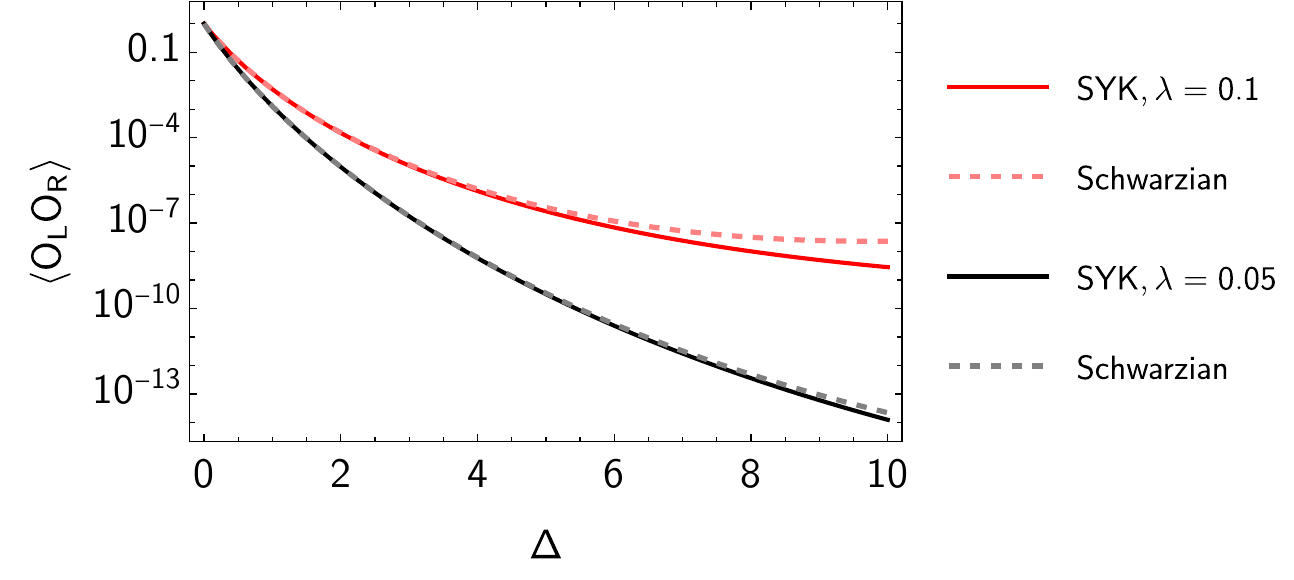}
\caption{Zero temperature 2-pt function in double scaled SYK as a function of $\Delta$. We are considering the fixed charge $j_R = 0$ ensemble. As $\lambda = 2p^2/N$ tends to zero, we recover the $\nn = 2$ super-Schwarzian predictions. }
\label{figg:2ptSch}
\end{figure}

The evaluation of the two-point function follows similar steps to that of the norm computation. Leaving the details to appendix \ref{app:twoPt}, we find that 
\begin{align}
\bra{\Psi,j} \q^{2\Delta \hat{n}} \ket{\Psi,j} 
= \frac{(\q^{2+4\Delta} ; \q^2)_\infty}
{(\q^{2+2\Delta} ; \q^2)_\infty^2  (\q^{1\pm2j_R+2\Delta} ; \q^2)_\infty}  
.
\end{align}
Normalizing the ground states and rewriting the result in terms of q-Gamma function $\Gamma_{q^2} (x) = (1-q^2)^{1-x} \frac{(q^2;q^2)_\infty}{(q^{2x};q^2)_\infty}$, we obtain 
\begin{align}
\frac{\bra{\Psi,j} \q^{2\Delta n} \ket{\Psi,j}}{\braket{\Psi,j}} 
= \frac{(1-q^2)^{2\Delta}}{\Gammaq(\hf \pm j_R)} 
\frac{\Gammaq(\Delta+1)^2 \Gammaq(\hf \pm j_R + \Delta)}{\Gammaq(2\Delta+1)} .
\end{align}
This form of the result can be immediately compared to the result obtained in $\mathcal{N}=2$ super Schwarzian theory. Taking the triple scaling limit $\q\to 1$ we find
\begin{align}
\frac{\bra{\Psi,j} \q^{2\Delta n} \ket{\Psi,j}}{\braket{\Psi,j}} 
\approx
   (2\lambda)^{2\Delta} \frac{\cos (\pi j_R)}{2\pi } \frac{\Delta \Gamma(\Delta)^2 \Gamma\left(\Delta+\frac{1}{2} \pm j_R\right)}{\Gamma(2 \Delta)}.
  \la{sch2predict}
\end{align}
Up to a renormalization, this is precisely what was obtained previously using $\mathcal{N}=2$ JT gravity techniques, see equation (85) of \cite{LongPaper}\footnote{In equation (85) of \cite{LongPaper}, there is an additional factor of $\cos \pi j/\hat{L}$ since they are normalizing the 2-pt function by the total number of BPS states, whereas we normalize by the number of ground states with charge $j_R$.
See also equation (59) of \cite{LongPaper}.}. See figure \ref{figg:2ptSch} for a detailed comparison. The answer for the two point function therefore takes a very simple form as a q-deformation of the super Schwarzian result. 
The above form also allows us to correctly identify the Schwarzian coupling $C = \alpha_S N =  (4 \lambda)\inv$, which indeed was predicted in Appendix H of \cite{Lin:2022rzw} by using the large-$p$ $G,\Sigma$ action.

\subsubsection*{Length of the wormhole}

As an application of these results, we can now compute the typical chord number in the ground state sector. This can be viewed as a discrete analogue of expectation value of the length of spatial wormhole in the bulk. 
By simply taking the derivative with respect to the scaling dimension $\Delta$ we find
\begin{align}
\ev{\ell} = 2 \lambda \langle n \rangle_{j_R} 
&= -2  \partial_\Delta \log[\bra{\Psi,j} \q^{\Delta n} \ket{\Psi,j} ]|_{\Delta=0} 
\\
&= - \lb {2\log(1-\q^2) + \psi_{\q^2}\lp \hf -j_R\rp  + \psi_{\q^2}\lp \hf +j_R\rp } \rb .
\end{align}
where $\psi_{q} (z) \equiv \frac{\partial_z \Gamma_q (z)}{ \Gamma_q (z)}$ is the q-Digamma function. 

\begin{figure}
\centering
\includegraphics[width=0.7\columnwidth]{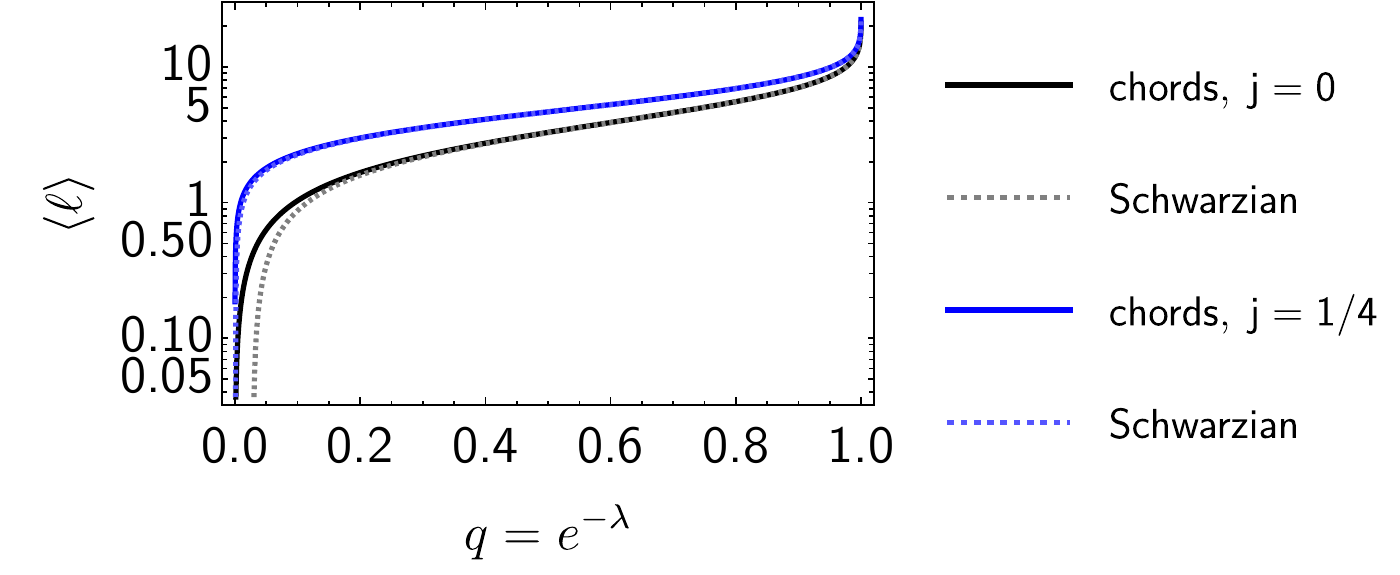}
\caption{Length as computed by the chords and the Schwarzian prediction. As $\q\to 1$ we recover the $\nn = 2$ super-Schwarzian predictions. }
\label{fig:length}
\end{figure}

Note that this length is finite away from the JT limit (see figure \ref{fig:length}). 
For $q\rightarrow 1$, we obtain that the expectation value for the renormalized length $\tilde{\ell} = \ell + 2\log(2\lambda)$ gives 
\be 
\label{eq:schLenPredict}
\langle \tilde{\ell} \rangle = -\psi(\hf -j_R) - \psi(\hf+j_R) ,
\ee
which is finite and matches the super Schwarzian prediction of \cite{LongPaper}. (Note that in the Schwarzian limit, the renormalized length of the wormhole is finite, but at finite $\lambda$ the ``bare'' length $\ell$ is finite.)
To go beyond the Schwarzian, we can consider the small $\lambda$ expansion.
Taking $j_R = 0$ for simplicity, %
\eqn{%
 \log(1-q^2) +  \psi_{q^2}\left(\hf \right) = -\frac{1}{2\lambda} \sum_{n=0}^\infty \frac{q^{2n+1}}{1-q^{2n+1}} =  \log (\lambda/2)    -\gamma_E  + 2 \sum_{g \text{ odd} }^\infty \frac{2^{g}-1}{g!} 
 \lambda^{g+1} \zeta^2(-g)  , }
Here $\zeta$ is the Riemann Zeta function and the sum is over positive odd integers $g=1,3,5, \cdots$. 
The non-perturbative corrections that are present in the q-Pochammer symbols cancel, so naively the renormalized length does not have non-perturbative corrections in $\lambda$. This seems surprising since the length wavefunction has non-perturbative corrections; however, one should be cautious in interpreting this statement since the sum is only asymptotic. It would be interesting to study the perturbative (and potentially non-perturbative) corrections to the length using the $G-\Sigma$ formalism, where $\ev{\ell}$ should be related to the 2-sided Liouville variable $g_{LR}$, see Appendix \ref{Liouville}.

\subsubsection*{2-point function of charged BPS operators}

Lastly, we can compute a two point function of a charged BPS operators with the $U(1)_{\mathrm{R}}$ charge $2\Delta$, 
using \nref{vbv2}.
${(\tilde\beta^\dag )}^{2\Delta_\O} q^{\hf \Delta_\O (n_\O + n_\X - j )} \equiv {(\tilde\beta^\dag )}^{4\Delta} q^{2 \Delta n} q^{-\Delta j} $. 
Such correlation functions can be computed by inserting a factor of \nref{vbv2} into \eqref{eq:norm}. Denoting $ j' =  -2j_R'$ and $j = -2j_R$, the normalized two-point function (for details see appendix
\ref{app:BPStwo_pt_function}) is
\begin{align}
 q^{-\Delta j } \frac{\bra{\Psi,j'} q^{2\Delta n}  (\tilde\beta^\dagger)^{4\Delta} \ket{\Psi,j}}{\braket{\Psi, j'}^{1/2} \braket{\Psi, j}^{1/2} }
&=  
q^{2\Delta j_R }
\delta_{j_R',j_R-2\Delta} \, 
\sqrt{\frac{(q^{1-2j_R};q^2)_\infty}
{(q^{1+2j_R};q^2)_\infty} 
\frac{(q^{1+2j'_R};q^2)_\infty}
{(q^{1-2j'_R};q^2)_\infty}} 
\\ 
&= 
q^{2\Delta j_R }
\delta_{j_R',j_R-2\Delta} \,  (1-q^2)^{j_R - j'_R} 
\sqrt{\frac{\Gammaq(\hf+j_R)}{\Gammaq(\hf-j_R)}
\frac{\Gammaq(\hf-j'_R)}{\Gammaq(\hf+j'_R)}}
.
\end{align}
Setting $j_R ' = j_R - 2\Delta$ and taking the triple scaling limit, this reduces to 
\be 
\frac{\bra{\Psi,j'}\V_L \vb_R \ket{\Psi,j}}{\braket{\Psi, j'}^{1/2} \braket{\Psi, j}^{1/2} } 
\approx 
q^{2\Delta j_R }
\, 
(2\lambda)^{2\Delta}
\sqrt{\frac{\cos(\pi j_R) \cos(\pi j'_R)}{\pi^2}} \Gamma(\hf + j_R) 
\Gamma(\hf - j'_R) , 
\ee
which is again consistent with the super-Schwarzian results \cite{LongPaper}. %

\pagebreak

\section{Wormholes with matter chords \la{sec:matter} }
In this section, we go beyond the empty wormhole and consider states with matter chords. 
Of course, if we are interested in the 2-pt function of matter operators, or more generally, any uncrossed $n$-pt function, we can use the Hilbert space of the empty wormhole to do the computation, as in the previous section. However, to consider crossed (out-of-time-order) correlators, one must go beyond the empty wormholes. 
One can also say that we are now considering ``vertical'' matter chords instead of just the ``horizontal'' chords in the previous section:
\begin{align}
\label{verticalChordDiagram}
\begin{tikzpicture}[scale=0.7, baseline={([yshift=-0.1cm]current bounding box.center)}, decoration={markings, mark=at position 0.2 with {\arrow{>}} }]
	\draw[thick]  (0,0) circle (1.6);
\begin{scope}[thick, decoration={markings, mark=at position 0.2 with {\arrow{>}} }]	
        \draw[postaction=decorate]  (0.93069294,1.30146481) -- (-1.13,-1.13)  ;
	\draw[postaction=decorate] (1.55025987,-0.39584) --(-1.05,1.2) ;
	\draw[postaction=decorate] (-1.58,-0.2) -- (1.13,-1.13) ;
	\draw[blue] (0,-1.6) -- (0,1.6) ; %
 \end{scope} 
	\draw[fill,black] (-1.58,-0.2) circle (0.1);
	\draw[fill,black] (-1.05,1.2) circle (0.1);
	\draw[fill,black] (1.55025987,-0.39584) circle (0.1);
	\draw[fill,black] (0.93069294,1.30146481) circle (0.1);
	\draw[fill,black] (-1.13,-1.13) circle (0.1);
	\draw[fill,black] (1.13,-1.13) circle (0.1);
	\draw[fill,black] (1.13,-1.13) circle (0.1);
	\draw[fill, blue] (0,-1.6) circle (0.1);
	\draw[fill, blue] (0,1.6) circle (0.1);
        \draw[dashed] (-1.6,0) -- (1.6,0);
        \draw[fill,gray] (-1.6,0) circle (0.1);
	\draw[fill,gray] (1.6,0) circle (0.1);
\end{tikzpicture} 
\qquad \RA \qquad 
\begin{tikzpicture}[scale=0.7, baseline={([yshift=-0.5cm]current bounding box.center)}, decoration={markings, mark=at position 0.2 with {\arrow{>}} }]
	\draw[thick]  (0,0) circle (1.6);
\begin{scope}[thick, decoration={markings, mark=at position 0.2 with {\arrow{>}} }]	
        \draw[postaction=decorate]  (0.93069294,1.30146481) -- (-1.13,-1.13)  ;
	\draw[postaction=decorate] (1.55025987,-0.39584) --(-1.05,1.2) ;
	\draw[postaction=decorate] (-1.58,-0.2) -- (1.13,-1.13) ;
	\draw[blue] (0,-1.6) -- (0,1.6) ; %
 \end{scope} 
	\draw[fill,black] (-1.58,-0.2) circle (0.1);
	\draw[fill,black] (-1.05,1.2) circle (0.1);
	\draw[fill,black] (1.55025987,-0.39584) circle (0.1);
	\draw[fill,black] (0.93069294,1.30146481) circle (0.1);
	\draw[fill,black] (-1.13,-1.13) circle (0.1);
	\draw[fill,black] (1.13,-1.13) circle (0.1);
	\draw[fill,black] (1.13,-1.13) circle (0.1);
	\draw[fill, blue] (0,-1.6) circle (0.1);
	\draw[fill, blue] (0,1.6) circle (0.1);
        \draw[dashed] (-1.6,0) -- (1.6,0);
        \draw[fill,white] (-1.7,0.1) rectangle ++(3.35,2.6); 
\end{tikzpicture} 
\qquad \RA \qquad 
\begin{tikzpicture}[scale=0.7, baseline={([yshift=-0.1cm]current bounding box.center)}, decoration={markings, mark=at position 0.2 with {\arrow{>}} }]
	\draw[thick]  (0,0) circle (1.6);
\begin{scope}[thick, decoration={markings, mark=at position 0.2 with {\arrow{>}} }]	
        \draw[postaction=decorate]  (0.93069294,1.30146481) -- (-1.13,-1.13)  ;
	\draw[postaction=decorate] (1.55025987,-0.39584) --(-1.05,1.2) ;
	\draw[postaction=decorate] (-1.58,-0.2) -- (1.13,-1.13) ;
	\draw[blue] (0,-1.6) -- (0,1.6) ; %
        \draw[postaction=decorate, red ] (-1.6,0) -- (1.6,0);
 \end{scope} 
	\draw[fill,black] (-1.58,-0.2) circle (0.1);
	\draw[fill,black] (-1.05,1.2) circle (0.1);
	\draw[fill,black] (1.55025987,-0.39584) circle (0.1);
	\draw[fill,black] (0.93069294,1.30146481) circle (0.1);
	\draw[fill,black] (-1.13,-1.13) circle (0.1);
	\draw[fill,black] (1.13,-1.13) circle (0.1);
	\draw[fill,black] (1.13,-1.13) circle (0.1);
	\draw[fill, blue] (0,-1.6) circle (0.1);
	\draw[fill, blue] (0,1.6) circle (0.1);
        \draw[fill,red] (-1.6,0) circle (0.1);
	\draw[fill,red] (1.6,0) circle (0.1);
\end{tikzpicture} 
\end{align}
In the above expression \nref{verticalChordDiagram}, we start with a diagram which when cut gives a state with a blue $\W$ chord in the middle of the wormhole. If we consider acting on this state with a ``horizontal'' chord such as $\V_L \vb_R$, we get a crossed or out-of-time-ordered correlator.
More conceptually, 2-sided wormholes with matter are an interesting class of states that can still preserve supersymmetry; understanding such SUSY wormholes in the chord formalism seems like an interesting direction.

\subsection{Supercharges for wormholes with matter}

First let us consider a non-BPS operator $\W$. A general 1-particle state can be made by concatenating two 0-particle states, while putting the operator $\W$ in between (see the coproduct discussion in \cite{newpaper}). Such states are labeled by $n_L, n_R$ and 4 qubits, e.g.:
\vspace{-0.2cm}
\begin{align}
    \ket{n_L,n_R,\X\O, \X \O}_\W &= |\overbrace{\X\O \X\O  \cdots \X\O}^{n_L \text{ pairs of } \X\O}  \W  \overbrace{\X \O\X\O \cdots \X\O}^{n_R \text{ pairs of } \X\O} \,  \rangle, \la{wstate} \\
    \ket{n_L,n_R,\O\O, \X \X}_\W &= |\overbrace{\O\X\O\X  \cdots \O\X}^{n_L \text{ pairs of } \O\X}  \O \W  \overbrace{\X \O\X\O \cdots \X\O}^{n_R \text{ pairs of } \X\O}\X   \rangle,
\end{align}
The first string (``XO'' in the first example and ``OO'' in the second example) labels the first and last character of the string that appears before the $\W$, whereas the second string (``XO'' in the first example, ``XX'' in the second) labels the first and last characters of the string that immediately follows the $\W$.

We may think of a state such as \nref{wstate} as Einstein-Rosen bridges with some matter particle $\W$ somewhere along the throat. %
Acting on this Hilbert space are some operators that can add or delete $X$'s. We define $\text{del X}_{LL}$ to delete an $\X$ from the beginning (the leftmost entry) of the string (or return 0 if the first entry is an $\O$). Similarly, we can define $\text{del X}_{ij}$, which act as %
\begin{align}
   \text{del X}_{LL} |\X\O  \cdots \X  \W  {\X \O \cdots \X} \,  \rangle &= |\O  \, \cdots\, \O \X  \W  {\X \O \cdots \O \X} \,  \rangle, \la{delXwstate1} \\
   \text{del X}_{LR} |\X\O  \cdots \X  \W  {\X \O \cdots \X} \,  \rangle &= |\X\O \, \cdots \, \O \W  {\X \O \cdots \O\X} \,  \rangle, \la{delXwstate2} \\
   \text{del X}_{RL} |\X\O  \cdots \X  \W  {\X \O \cdots \X} \,  \rangle &= |\X\O  \cdots \O \X  \W  { \O \, \cdots \, \O\X} \,  \rangle, \la{delXwstate3} \\
   \text{del X}_{RR} |\X\O  \cdots \X  \W  {\X \O \cdots \X} \,  \rangle &= |\X\O  \cdots \O\X  \W  {\X \O \, \cdots \, \O} \,  \rangle. \la{delXwstate4} 
\end{align}
If there is no $X$ in the location to be deleted, the operator returns zero. We can similarly define $\text{del O}_{ij}$. For the operators that add, we will only need the operators which add an $\X$ or $\O$ in the left-most or right-most entry $\text{add O}_{ii}, \text{add X}_{ii}$. These operators also annihilate any state that already has an $\X$ ($\O$) in the corresponding location, e.g. $\text{add X}_{RR} \ket{ \O\X \cdots \O \W \O \X \cdots \X} = 0$.

To write the supercharges with a matter chord, recall the expressions for the supercharges with 0 matter particles \nref{supercharges}.
Focusing on the ``delete'' terms, we had $Q_R \supset  ( A \, \text{del X}_{\lt} + B \, \text{del X}_{R} ) \beta   $ and $(-1)^F Q_L \supset ( A \, \text{del X}_{\lt} + B \, \text{del X}_{R} ) \beta^\dagger $ where $A$ and $B$ are given in \nref{abdef}.  
Let's first assume that the matter chord is close to the identity (e.g. $(-1)^{F_m} = 1$ and $\Delta_\O, \Delta_\X \ll 1$.) 
We expect that $Q_R$ should take the form
\begin{align}
Q_R &\sim  A_R ( A_L \, \text{del X}_{\lt\lt} + B_L \, \text{del X}_{\lt\rt} ) +B_L  ( A_R \, \text{del X}_{\rt\lt} +  B_R \, \text{del X}_{\rt\rt}) \beta + q^{\tilde{m} /2}\, \text{add O}_{RR}. \la{babyCoproduct}
\end{align}
Here $A_{L/R} = (-1)^{F_{L/R}} q^{(n_\tot)_{L/R} / 2} $ and $B_{L/R} = q^{(n_{\X} - n_\O )_{L/R}/2} $.
Now it is easy to write down the more general case:
\begin{align} %
Q_R = & \, (-1)^{F_R + F_m} q^{\hf (\Delta_\X + \Delta_\O + (n_{\X}  + n_{\O })_R  ) }  ( A_L \text{del X}_{\lt\lt} + B_L \text{del X}_{\lt\rt} ) \beta \la{gentleCoproduct} \\
 &+ q^{\hf(\Delta_\X- \Delta_\O +(n_{\X} - n_{\O})_L )} ( A_R \text{del X}_{\rt\lt} +  B_R \text{del X}_{\rt\rt}) \beta + q^{\tilde{m} /2}\, \text{add O}_{RR} \notag \\
 \bq_R = & \, (-1)^{F_R + F_m} q^{\hf (\Delta_\X + \Delta_\O + (n_{\X}  + n_{\O})_R ) }  ( \bar{A}_L \text{del O}_{\lt\lt} + \bar{B}_L \text{del O}_{\lt\rt} )\beta^\dagger \la{gentleCoproduct2} \\
 &+ q^{\hf(\Delta_\O- \Delta_\X +(n_{\O} - n_{\X})_L )} (\bar{A}_R \text{del O}_{\rt\lt} + \bar{B}_R \text{del O}_{\rt\rt})\beta^\dagger+ q^{-\tilde{m} /2}\, \text{add X}_{RR} \notag \\
(-1)^{F_\text{tot}} Q_L = &\, (-1)^{F_L + F_m} q^{\hf (\Delta_\X + \Delta_\O + (n_{\X}  + n_{\O })_L ) }  ( A_R \text{del X}_{RR} + B_R \text{del X}_{RL} ) \beta^\dagger \\
 &+ q^{\hf(\Delta_{\X}- \Delta_{\O} +(n_{\X} - n_{\O} )_R )} (\bar{A}_L \text{del X}_{LR} +  \bar{B}_L \text{del X}_{LL})\beta^\dagger + q^{\tilde{m} /2}\, \text{add O}_{LL}  \notag \\
\bq_L (-1)^{F_\text{tot}}   = &\, (-1)^{F_L + F_m} q^{\hf (\Delta_\X + \Delta_\O + (n_{\X}  + n_{\O })_L ) }  (\bar{A}_R \text{del O}_{RR} + \bar{B}_R \text{del O}_{RL} ) \beta \\
 &+ q^{\hf(\Delta_\O- \Delta_\X +(n_{\O} - n_{\X} )_R )} ( \bar{A}_L \text{del O}_{LR} +  \bar{B}_L \text{del O}_{LL})\beta+ q^{-\tilde{m} /2}\, \text{add X}_{LL}  \notag 
\end{align}
A more elegant presentation of these supercharges will be discussed shortly; we wrote these expressions for concreteness. %
Here $(-1)^{F_\text{tot}}  = (-1)^{F_L + F_m+F_R}$. We can also define $(n_\O - n_\X)_\text{tot} = (n_\O - n_\X)_L + (n_\O - n_\X)_R +\Delta_\O - \Delta_\X $. We also have the expressions for the charges:
\begin{align}
    J_R &= \frac12\left[ (n_\O  - n_\X)_\text{tot}  + m\right] - \frac{1}{\lambda} \pd_m \\
    J_L &= \frac12\left[ (n_\O - n_\X)_\text{tot} -m\right] + \frac{1}{\lambda} \pd_m 
\end{align}
Using these expressions, we explicitly verified the SUSY algebra \nref{Malg1}-\nref{Malg3} by writing the explicit action of the supercharges on 16-component wavefunctions in {\it Mathematica} (see the notebook in the Supplementary Materials of the JHEP version of this paper). Whereas for empty wormholes, we had $\{Q_L ,\bq_L\} = \{Q_R, \bq_R\}$; now the supercharges define different Hamiltonians $H_L$ and $H_R$ that mutually commute:
\begin{align}
    \{Q_i,Q_j\} &= \{ \bq_i, \bq_j \} = 0, \la{Malg1}\\
    \{Q_i, \bq_j\} &= \delta_{ij} H_i,\la{Malg2}\\
    [H_i,H_j] &= 0,\la{Malg3}%
\end{align}
Here $i,j$ run over $L$ and $R$. We also continue to have the same $U(1)_\mathrm{R}$ relations \nref{u1r1} and \nref{u1r2}.

Note that using these expressions, one can in principle compute the OTOC 4-pt function. In particular, one can derive the action of a horizontal chord, etc., using similar considerations to the derivation for the empty wormhole. For example, the neutral operator gives
\eqn{\W_L \W_R \sim q^{\Delta(n_{\O,\tot} + n_{\X,\tot}}).}
Evaluating this operator on the state obtained by solving the equations $Q_R \ket{\Psi,j} = \qb_R \ket{\Psi,j} = Q_L \ket{\Psi,j} = \qb_L \ket{\Psi,j} =0$ would give the zero temperature OTOC. It would also be interesting to analyze the triple scaling limit of \nref{gentleCoproduct}.
For fixed boundary $U(1)$ charges, we expect at most one zero energy bound state for each chord primary if the R-charge of the primary $[J_R, \W] = j_W \W$ is such that $|j_W| < 1$ and the charge of the state is in the appropriate range. %

Equipped with the Hilbert space of wormholes with matter, one can write down various generalizations of \nref{superBowlAd}. For example,
\eqn{ 
\text{probability(length of SUSY wormhole $=0, \W$)} \; = \; \frac{ \Tr (\Pi_{0,j_R} \W \Pi_{0,j_R} \W)} { \dim \mathcal{H}_{j_R} }  \la{superAdMatter}
.}
To compute the LHS, we use the Born rule applied to the superchord Hilbert space, e.g., we compute the overlap between a SUSY wormhole with one matter particle of type $\W$, and the wormhole state with minimal length, e.g., a state with no $\O$'s or $\X$'s. (Both wormhole states have $J_L + J_R = 0$ and $J_R  = j_R$; the SUSY wormhole wavefunction would again be determined by finding the state annihilated by all four supercharges in \nref{gentleCoproduct}.) On the RHS, $\Pi_{0,j_R} = e^{- \infty H} \Pi_{j_R}$ is a projector onto the zero energy subsector with the appropriate $U(1)_\R$ charge.

If we replaced the LHS of \nref{superAdMatter} with some matrix elements of chord operators that involved $n_\O$ and $n_\X$, we would obtain a computation that involved both a horizontal chord associated with the chord operator and a vertical chord associated with the matter that lives in the particular state. Thus this would modify the RHS of \nref{superAdMatter} to the 4-pt OTOC, as depicted schematically in \nref{verticalChordDiagram}.

\subsection{Superchord algebra}
\def\frakx{\texttt{X}}
\def\frako{\texttt{O}}
The superchord algebra is generated by the supercharges $Q_L, Q_R, \bq_L, \bq_R$ as well as two other operators $n_\O$ and $n_\X$. Microscopically, $n_\O$ and $n_\X$ these are related to ``operator size'', see \cite{Roberts:2018mnp, Qi:2018bje, Lin:2022rbf}. For Dirac fermions, the size operator is defined as %
\begin{align}
 \text{size} &= \frac{1}{2} \sum_{\alpha = 1}^N  1+  \i  \psi_\alpha^L \psib_\alpha^R + \i \psib_\alpha^L \psi_\alpha^R %
\end{align}
In a theory with a $U(1)$ symmetry, we can then define two kinds of sizes:
\begin{align}
    \text{size}_\O = \text{size} - J_L + J_R\\
    \text{size}_\X = \text{size} + J_L - J_R
\end{align}
Here $\text{size}_\O$ counts the number of $\psi$ fermions in a given operator, whereas $\text{size}_\X$ counts the number of anti-fermions $\psib$. Then,
\begin{align}
   \frac{\text{size}_\O }{p} = n_{\O }, \qquad 
   \frac{\text{size}_\X }{p} = n_{\X }
.
\end{align}

To write the algebra, we split the supercharges into creation and annihilation terms:
\begin{align} \la{superchargesSplit}
    Q_R &= \frakx_R \beta^\dagger + q^{m/2} \frako_R^\dag\\
\qb_R &= \frako_R \beta + q^{-m/2} \frakx_R^\dag\\
Q_L &= \frakx_L \beta + q^{-m/2} \frako_L^\dag \\
\qb_L &= \frako_L \beta^\dag + q^{m/2} \frakx_L^\dag  \la{superchargesSplit4}
\end{align}
These operators satisfy an algebra which we call the ``superchord algebra'':
\begin{align}
    &[n_\X, \frakx_{L/R}] = -\frakx_{L/R}, \quad [n_\X, \frakx_{L/R}^\dag] = \frakx_{L/R}^\dag \la{chordb}\\
    &[n_\O, \frako_{L/R}] = -\frako_{L/R},\quad 
    [n_\O, \frako_{L/R}^\dag] = \frako_{L/R}^\dag \\
    &\{ \frakx_{L/R} , \frakx_{L/R}^\dag\}_{\sqrt{q}} =  q^{\hf (n_\X - n_\O) } \\
    &\{ \frako_{L/R} , \frako_{L/R}^\dag\}_{\sqrt{q}} = q^{\hf (n_\O - n_\X)} \\
    &\{ \frakx_{L/R}, \frakx_{R/L}^\dag \}_{\sqrt{q}} = - \{ \frako_{L/R}, \frako_{R/L}^\dag\} _{\sqrt{q}} = q^{\hf n_\text{tot} } \la{chorde}
\end{align}
The rest of the anti-commutators are zero:
    \begin{align} 
    &\{ \frakx_L, \frakx_R\} = \{ \frako_L, \frako_R\} = \{ \frakx_L^\dag, \frakx_R^\dag \} =\{ \frako_L^\dag, \frako_R^\dag\} =  0 \la{triv1}\\
    &\{ \frakx_L, \frako_R\} = \{ \frako_L, \frakx_R \} = \{ \frakx_L^\dag, \frako_R^\dag\} = \{ \frako_L^\dag, \frakx_R^\dag \} = 0\la{triv2}
    \end{align}
    In addition, $(-1)^F$ anticommutes with all of the $\X$ and $\O$ operators and commutes with the $n$ operators.
Using the explicit expressions for the supercharges, we checked using {\it Mathematica} that the above algebra is satisfied. In the case of the 1-particle expressions, we used $n_\X = (n_{\X,L} + n_{\X,R} + \Delta_\X)$ and $n_\O = n_{\O,L} + n_{\O, R} + \Delta_\O)$. 
We also have the relations for the $U(1)$ rotor:
\begin{align}
    &[\beta, q^{m/2}]_{\sqrt{q}} =[\beta^\dag, q^{-m/2}]_{\sqrt{q}} = 0, \quad \beta \beta^\dag = 1
\end{align}
These relations undoubtedly have many applications, see \cite{newpaper}. As a simple one, we can use them to prove that \nref{Malg1}-\nref{Malg3} is satisfied.

An interesting subalgebra %
is generated by the 8 elements 
\begin{align}
    \chi_{ij} = \frakx_i^\dag \frakx_j, \quad \phi_{ij} = \frako_i^\dag \frako_j,\\
    [n_\X, \chi_{ij}] = [n_\X, \phi_{ij} ]= [n_\O, \chi_{ij}] = [n_\O, \phi_{ij} ]=  0
\end{align}
The subalgebra satisfies a somewhat complicated set of relations, see Appendix \ref{app:subalg}.

It would be interesting to study this subalgebra further, in particular in the triple scaling limit where we expect it to be related to the bulk supersymmetries $\mathfrak{osp}(2|2)$ or $\mathfrak{su}(1,1|1)$ \cite{Lin:2019qwu}. Furthermore, the full superchord algebra should contract to the JT superalgebra; so far only the bosonic JT algebra has been discussed in the literature \cite{Harlow:2021dfp, Lin:2022rbf}.

We expect that this subalgebra has a Casimir that can be promoted to a full Casimir of the superchord algebra. 
Such a Casimir would commute $Q_L, \qb_L, Q_R, \qb_R$, and thus it could be used to distinguish different supersymmetric wormholes. 
In particular, such an operator would be closely related to the bulk ``matter energy'' which can be non-zero even though the ADM mass of such wormholes is exactly zero (by supersymmetry). It therefore seems related to the emergence of bulk time, even though there is no boundary time \cite{LongPaper}.

This superchord algebra is likely to provide strong constraints on any proposed gauge-theoretic description of the chord theory, see \cite{Berkooz:2022mfk} and \cite{Blommaert:2023opb} for some recent discussions in the non-supersymmetric case.

\subsection{Superchord coproduct}
The above algebra \nref{chordb}-\nref{triv2} is actually a bi-algebra with a coproduct $D$, which has three parameters $\Delta_\X, \Delta_\O$ and the sign $(-1)^{F_m}$.
The coproduct $D$ is a coassociative algebra homomorphism from $\mathcal{A} \to \mathcal{A} \otimes \mathcal{A}$. It is a technical tool that allows one to build multi-particle representations of the chord algebra \cite{newpaper}, starting from 1-particle representations.
\begin{align}
D((-1)^F) &= (-1)^{F_m} (-1)^F \otimes (-1)^F\\
D(n_\X) &= n_\X \otimes 1 + 1 \otimes n_\X + \Delta_\X ( 1 \otimes 1)\\
D(n_\O) &= n_\O \otimes 1 + 1 \otimes n_\O + \Delta_\O (1 \otimes 1)\\
D(\frakx_R^\dag) &= 1 \otimes \frakx_R^\dag \\
D(\frako_R^\dag) &= 1 \otimes \frako_R^\dag \\
D(\frakx_L^\dag) &= (-1)^{F_m} \frakx_L^\dag  \otimes (-1)^F \\
D(\frako_L^\dag) &= (-1)^{F_m} \frako_L^\dag \otimes (-1)^F \\
D(\frakx_R) &= \frakx_R \otimes (-1)^{F_m + F}q^{(\Delta_\X +\Delta_\O + n_\X + n_\O) /2}  +  q^{(\Delta_\X - \Delta_\O + n_\X - n_\O)/2}     \otimes \frakx_R \\
D(\frako_R) &= \frako_R \otimes (-1)^{F_m + F}q^{(\Delta_\X +\Delta_\O + n_\X + n_\O) /2}  +  q^{(\Delta_\O - \Delta_\X + n_\O - n_\X)/2}     \otimes \frakx_R \\
D(\frakx_L) &= q^{(\Delta_\X +\Delta_\O + n_\X + n_\O) /2}  \otimes \frakx_L + \frakx_L \otimes (-1)^{F_m+F} q^{(\Delta_\X - \Delta_\O + n_\X - n_\O)/2}\\
D(\frako_L) &= q^{(\Delta_\X +\Delta_\O + n_\X + n_\O) /2}  \otimes \frako_L + \frako_L \otimes (-1)^{F_m + F}q^{(\Delta_\O - \Delta_\X + n_\O - n_\X)/2}
\end{align}
One can check that the coproduct $D$ preserves the chord superalgebra. As a sample computation:
\begin{align}
    &\{ D(\frakx_R) , D(\frakx_R^\dag ) \}_{\sqrt{q}}  = D(\frakx_R) D(\frakx_R^\dag) + \sqrt{q} D(\frakx_R^\dag) D(\frakx_R)\\
    &\; = \frakx_R \otimes \{ (-1)^{F_m + F}q^{(\Delta_\X +\Delta_\O + n_\X + n_\O) /2} , \frakx_R^\dag \}_{\sqrt{q} }+  q^{(\Delta_\X - \Delta_\O + n_\X - n_\O)/2}     \otimes \{\frakx_R, \frakx_R^\dag\}_{\sqrt{q}}\\
    &\; = D\big(q^{\hf(n_{\X} - n_{\O})}\big)
\end{align}
In writing the last line, we used that $\{(-1)^F q^{n_\X / 2}, \frakx^\dag \}_{\sqrt{q}} = (-1)^F [q^{n_\X / 2} , \frakx^\dag ]_{\sqrt{q} }= 0$ according to \nref{chordb}.

In analogy with the $\mathcal{N}= 0$ case \cite{newpaper}, we expect that the empty wormhole gives a ``short'' irrep of the superchord algebra and that the 1-particle expressions \ref{gentleCoproduct} define a generic irrep of the superchord algebra. It would be interesting to analyze this further.

As a simple application of the coproduct, we can apply the coproduct on the expressions \nref{superchargesSplit}-\nref{superchargesSplit4} to recursively obtain expressions for the supercharges acting on multi-particle states. In other words, the coproduct saves us the hassle of working out the chord combinatorics for wormholes with multiple matter chords; if we know the form of the supercharges acting on wormholes with $m$ matter chords, the coproduct immediately returns the supercharges for $m+1$ matter chords.

For a chiral/BPS operator, recall that $[Q,\V] = 0$ so a chord state like $Q \V Q \ket{\Omega} = Q^2 \V \ket{\Omega} = 0$ and similarly $\qb \vb \qb \ket\Omega = 0$. This implies that we only need to consider chord states where $\V$ is sandwiched between at least one $\X$, e.g., $\X \V \X, \X \V \O, \O \V \X$. Hence the BPS operators generate a shorter representation of the superchord algebra than a generic operator.

\section{Discussion \la{discussion}}
\subsection{Bulk-to-boundary  map \label{bbmap}}
In \cite{Lin:2022rbf} an algorithm for constructing arbitrary states in the chord basis in terms of the microscopic states was given. This can be viewed as providing a bulk-to-boundary map that is exact at the level of the disk (e.g. in the triple scaling limit it accounts for the Schwarzian fluctuations). Without matter particles, the algorithm was particularly simple and involved a Gram-Schmidt orthogonalization of the states $\ket{\Omega}, H \ket{\Omega}, H^2 \ket{\Omega}, \cdots$. In this context, the chord number can then be identified with the Krylov complexity of the 2-sided state \cite{Lin:2022rbf}. For some recent discussions of Krylov complexity, see \cite{Rabinovici:2023yex,Balasubramanian:2022tpr}.

In the $\mathcal{N} = 2$ context, a similar algorithm exists. Let us explain the case without matter. 
\begin{enumerate}
    \item Write down the states $\ket{n, ab, j}$ where $a,b \in \{ \O, \X\}$ using the string notation \nref{string}, e.g., $\ket{n,\X\O,j} =  |\overbrace{\X\O \X\O  \cdots \X\O}^{n \text{ pairs of } \X\O} , j \rangle$. %
    \item Consider the closely related states obtained by replacing $\O$ with $Q$ and $\X$ with $\qb$ acting on the maximally entangled states, e.g., 
    \eqn{\ket{n,\X\O,j} \to \overbrace{\qb_R Q_R \qb_R Q_R  \cdots \qb_R Q_R }^{n \text{ pairs of } \qb Q} \ket{\Omega,j}.}
    For the states with an odd number of supercharges, one also needs to adjust the value of $j$, e.g., \eqn{\ket{n,\X\X,j} \to \overbrace{\qb_R Q_R \qb_R Q_R  \cdots \qb_R Q_R }^{n \text{ pairs of } \qb Q} \ket{\Omega,j+1}.}
    The shift in the charge ensures that after the action of the supercharges, the resulting state has charge $j$.
    The states on the RHS can be viewed as states that live in the microscopic Hilbert space of 2 SYK models. Our goal is to obtain the states on the LHS.
    \item For each value of $j$, list the states in order of increasing $n$ and then perform a Gram-Schmidt orthogonalization. Only orthogonalize states with respect to other states of smaller total $n_\X + n_\O$.\footnote{E.g., do not orthogonalize a state $Q_R \qb_R Q_R \ket{\Omega,j+1}$ with respect to $\qb_R Q_R \qb_R \ket{\Omega, j-1}$.} Equivalently, one can use the explicit form of the supercharges \nref{jqrt}-\nref{jqltb} to determine what states to subtract off (the ``delete'' terms) in the supercharges. %
\end{enumerate}
Just to illustrate the procedure, let us work out the first few states. First the state with no chords is the maximally entangled state $\ket{\Omega,j}$.
Then the state with a single O or X chord is $Q_R\ket{\Omega,j+1} = \ket{\O,j}$, $\qb_R \ket{\Omega, j-1} = \ket{\X,j}$.
The first non-trivial states are then
\begin{align}
 \ket{\X\O,j} &= \qb_R Q_R \ket{\Omega,j} -q^{-\frac{j}{2}+\tfrac14 } \ket{\Omega,j} ,  \\
\ket{\O\X,j} &= Q_R \qb_R \ket{\Omega,j}- q^{+\frac{j}{2}+\tfrac14 }\ket{\Omega,j} , \\
\ket{\X\O\X,j} &= \qb_R \ket{\X\O,j-1} - q^{-\frac{j}{2}+\tfrac14 }\qb_R  \ket{\Omega,j-1} ,\\
\ket{\O\X\O,j} &= Q_R \ket{\O\X,j+1} - q^{+\frac{j}{2}+\tfrac14 }Q_R  \ket{\Omega,j+1}
.
\end{align}

As in the $\mathcal{N} = 0$ case, this Krylov/Gram-Schmidt procedure can be generalized to cases with arbitrary numbers of matter chords, although suitable caution must be taken to only orthogonalize states that should actually be orthogonal. Since the supercharges can only create or delete one chord, if we list states in order of increasing chord number, the supercharges are in tri-diagonal form. (For the case with matter particles, the coproduct construction implies that the supercharges will always have this form.) This tri-diagonal form is the basic reason why an orthogonalization procedure works in all cases.

At finite $\lambda$, we have seen that the bound state wavefunction corresponding to the supersymmetric wormhole has finite chord number. Furthermore, the wavefunction decays rapidly at large chord number, so that it is well approximated by truncating to states with some maximal number of chords. This implies that the empty supersymmetric wormhole is in some sense ``simple'', e.g., its Krylov complexity is bounded. Note that at $\lambda \gg 1$, the state is approximated by just a wormhole with a small number of chords, and hence its Krylov complexity is small. 
As an aside, note that if one wanted to describe a 1-sided extremal microstate in the SYK theory at finite $N$, one can in principle describe it as a complicated superposition of ``simple'' states; however, if the dimension of the ground state sector is almost as large as the $\dim \mathcal{H},$ then there will exist ground states where the superposition is much simpler.

\subsection{Comments on the boundary length/sign problem}

In \cite{LongPaper}, a simple puzzle was posed about the geometry of the disk in the $\mathcal{N}=2$ super JT theory in the zero temperature limit. %
 The puzzle is that according to the super-Schwarzian theory, the renormalized boundary length of the disk $\beta/\epsilon $ diverges as the temperature goes to zero, whereas the typical length between points on the boundary stays finite. It seems hard to find any non-backtracking\footnote{E.g. curves such that the angular coordinate is a monotonic function of the proper length.} curves on the Poincare disk that have arbitrary long length with this property.

Since the chord geometry gives a picture of sorts of the bulk, we may ask whether this geometry can be used to clarify the puzzle.
It will be useful to compare with the bosonic $\caln = 0$ theory. The idea is to view the partition function at some inverse $\beta$ as a sum over chord diagrams, and ask about some properties of a ``typical'' diagram that appears in the sum. Since
\eqn{Z(\beta) = \sum_k \frac{(-\beta)^k}{k!}\tr H^k , \la{sum} } 
we must sum over chord diagrams with $k$ boundary points. Note that there is a dramatic difference in the sum \nref{sum} between the bosonic case and the supersymmetric case. In the $\caln = 0$ model, all odd moments vanished and we can interpret the quantity $\beta^{2k} \tr (H^{2k}) /(2k)! $ as a measure for the boundary length. At large $\beta$, this essentially gives a Poisson distribution
\eqn{\text{probability(chord diagram has boundary length } k)\propto (\beta E_0)^{2k}/(2k)!
}
Hence the typical length in the $\nn = 0$ theory at low temperature is
\eqn{\text{boundary length} \equiv k \lambda \approx \hf \lambda \beta |E_0| \approx  \beta \mathcal{J} }
Here we have defined the boundary length with a factor of $\lambda$ since bulk lengths are defined with the same factor $\ell = \lambda n$.
Note that with this normalization, the fluctuations in the boundary length are suppressed as $\lambda \to 0$ in the low temperature limit, as expected from the Schwarzian theory (where the boundary length is fixed to be $\beta/\epsilon$ with no fluctuations tolerated.) This means that if we wish to study the ``typical'' bulk geometry of the low temperature $\mathcal{N}=0$ theory, we may simply consider the set of chord diagrams that contribute to $\tr H^{2k}$ where $k \sim \beta E_0/2$.

On the other hand, no such simple procedure is known to us in the $\mathcal{N}=2$ theory. The sum over chord diagrams is not positive, in part\footnote{Even if we consider the computation of $\tr H^k$ for fixed $k$, the chord rules \nref{enemies3}, \nref{enemies4} imply that certain diagrams get a negative weight.} because the odd moments do not vanish and thus we cannot neglect the $(-1)^k$. 
To illustrate the severity of the sign problem, imagine ignoring the $(-1)^k$. Then, like in the $\nn=0$ case, we would get a boundary length that is peaked around a value that grows linearly with $\beta$ at large $\beta$, in line with Schwarzian expectations. 
However, one can argue that moments of the bulk length $\ell$ of the 2-sided state $H^k \ket{\Omega}$ diverge as $k \to \infty$, despite the fact that the bulk lengths stay finite as $\beta \to \infty$.
Thus from this perspective, the fact that the bulk length stays finite comes from the detailed cancellations $(-1)^k$ between chord diagrams with similar boundary lengths.

\subsection{Comments on more general extremal black holes \la{sec:discussionGen}}
Let's now comment on microstate counting for more general extremal black holes, such as those that can be embedded in a higher dimensional AdS setup with a more traditional CFT dual.  Morally speaking, we still expect an equation like \nref{superBowlAd}, except now there are 2 imprecisions.
First, instead of $\ell \to 0$, we should really discuss the renormalized length $\tilde{\ell} \to -\infty$. Second, the fraction of states is really infinitesimal.\footnote{The fact that these issues didn't arise in double-scaled SYK, e.g., the fraction is finite and the bare length $=0$ (maximally entangled) state is part of the Hilbert space, is related to the statement that the algebra of one-sided observables in double scaled SYK is Type II$_1$ algebra.}

Actually, there is a simple modification of this equation that resolves these issues, which also follows from \nref{micro}.
One replaces the maximally entangled state $\ket\Omega$ with the thermofield double at some high temperature $1/\beta$: 
\eqn{
 |\braket{\text{BPS},j}{\TFD, \beta,j}|^2 = \frac{Z_{\text{BPS},j_R}}{Z_{j_R}(\beta) } \la{regulated} 
}
We can interpret the LHS of this equation as the probability that the BPS wormhole has the same geometry as a high-temperature wormhole. Such a wormhole will have a short throat compared to that of an extremal black hole, which has a long nearly-AdS$_2$ throat. In $\mathcal{N}= 2$ super JT gravity, the bound state wavefunction goes to zero at $\tilde \ell \to - \infty$ like $|\psi(\tilde\ell)|^2 \sim \exp(-4e^{-\tilde\ell/2})$. This decay of the wavefunction can be easily understood in the Liouville picture, since we are evaluating the $E=0$ wavefunction in the forbidden region with a Liouville potential $\propto e^{-\tilde\ell}$. On the other hand, at high temperatures, the Schwarzian mode is semi-classical and the boundary is a circle of radius $\ell_*/2$. On the Poincare disk, its renormalized circumference is therefore $\beta \propto e^{\tilde\ell/2}$. This reproduces the correct high temperature behavior of the ratio \nref{regulated}, which for $j_R=0$ goes like $ e^{- \pi^2/\beta}$. Of course, equation \nref{regulated} is actually precise for any value of $\beta$ in the $\nn =2$ Schwarzian theory, but the point is that regulating $\tilde\ell \to -\infty$ is closely related to regulating the infinite temperature partition function by introducing a small $\beta$, see also \cite{Penington:2023dql, Kolchmeyer:2023gwa}.  These comments also generalize to relations with higher-pt functions, such as \nref{superAdMatter}.

In some other supergravity theories, there are contributions to the BPS partition function from other supersymmetric geometries, such as ones with conical defects \cite{Iliesiu:2022kny,Boruch:2022tno,Turiaci:2023jfa,Boruch:2023trc}. We can view such conical defects\footnote{It would be interesting to understand if the non-perturbative corrections to the degeneracy can be viewed as a sum over saddles in the $G,\Sigma$ formulation. In the double scaled theory, the $G,\Sigma$ integral is an unconventional 2D super-Liouville theory. As a side remark, we do not believe that such saddles (if they exist) would correspond to conical defects as the $j_R$ dependence of such contributions does not match.} as modifying the probability that the supersymmetric wormhole is ``very short.'' %

Finally, let us give a cautionary remark: it is sometimes said that low-energy supergravity is smart enough to reproduce the exact microstate count for certain extremal black holes. While it is certainly remarkable that this is true in some examples, this seems to fail in the context of super SYK, since the super-Schwarzian theory alone does not seem to reproduce the exact degeneracy. 
In the case of double scaled SYK, one had to use the ``UV completion'' of the gravity theory in order to reproduce the non-perturbative corrections to the degeneracy. Indeed, the fact that the length is discrete in the chord theory and not continuous like in gravity seems to be closely related to the existence of these corrections. It would be interesting to find other examples of this phenomena in higher dimensions. 

\subsection*{Acknowledgments}
We thank Pawel Caputa, Luca Iliesiu, Sonja Klisch, Juan Maldacena, Henry Maxfield, Stephen Shenker, Douglas Stanford, and Zhenbin Yang for helpful discussions.
JB is supported by the NCN Sonata Bis 9 2019/34/E/ST2/00123 grant. JB wants to give special thanks to Douglas Stanford and the Stanford Institute for Theoretical Physics for generous hospitality and support during his visit at Stanford University, where this work was initiated.
HL is supported by a Bloch Fellowship. %
CY wants to give special thanks to Douglas Stanford and Shunyu Yao for helping her get through difficult times in life which coincided with when this work was initiated. 
\newpage

\appendix

\section{Review of microscopic ground state count \la{app:index} }

In the main text, we calculated the fraction of ground state $\hat{D}_j$ from the bulk perspective. In this Appendix, we look at our calculation of the fraction of ground state from the boundary perspective: the Fourier transform of the refined index. We find agreement between our calculation and existing results of \cite{Fu:2016vas}.

In this section, we review the refined index computation of \cite{Fu:2016vas} with our conventions. For simplicity, we focus on the case where $N$ is even. The supercharges commute with a discrete subgroup of $U(1)_{\mathrm{R}}$ symmetry given by $e^{2 \pi \i r J}$, where $r \in (0,1, \cdots p-1)$. This allows one to define a refined index, and evaluate it in the free theory
\begin{align}
W_r &= \operatorname{tr}\left[(-1)^F e^{2 \pi \i r J}\right] = 2^{-N} \left[e^{ \frac{ \pi \i r}{p} } -e^{-\frac{\pi \i r}{p}}\right]^N = (-1)^{N/2}\sin^N (\pi r/p)
\label{wr}
\end{align}
Here we use a convention where a single Dirac fermion has 2 states: a bosonic state with $J = + 1/p$ and a fermionic state with $J= -1/p$. Note that with this convention a $j_R = 0$ state in the Hilbert space is bosonic (fermionic) if $N/2$ is even (odd).
By assuming that the index is saturated, one can write 
\begin{align}
W_r &=   \sum_{j_R}  \hat{D}(j_R) \, e^{2\pi \i r j_R} (-1)^{j_R p + N/2} \\
&=  \sum_{j_R}  \hat{D}(j_R) \, e^{2\pi \i r j_R} e^{-\pi i j_R p + N/2}
\end{align}
Then performing an inverse discrete Fourier transform and relabeling the index gives %
\begin{align}
\hat{D}(j_R)&=\frac{e^{\pi i j_R p}}{p}\sum_{n=0}^{p-1}\left[\sin(\frac{\pi n}{p})\right]^N e^{-2\pi \i j_R n}\\
&=\frac{e^{\pi i j_R p}}{p}\sum_{m \text{ even, } m=2}^{2p-2}\left[\sin(\frac{\pi m}{2p})\right]^N e^{-\pi \i j_R m}\\
 &=\frac{1}{p}\sum_{m \text{ odd, } m=2-p}^{p-2}\left[\cos(\frac{\pi m}{2p})\right]^N e^{-\pi \i j_R m}
\end{align}
Hence we arrive at the result (see Appendix G of \cite{LongPaper}):
\begin{align}
\hat{D}(j_R)
&=\frac{1}{p} \sum_{l=0}^{(p-3) / 2} 2 \cos \left({\pi j_R(2 l+1)}\right)\left[ \cos \frac{\pi}{2 p }(2 l+1)\right]^N \label{riresult}
\end{align}
Written this way, $\hat{D}$ is manifestly positive; one can check that $\hat{D} 2^N$ is an integer. Note that for $p=3$ the ground state count exactly matches the Schwarzian prediction. However for $p>3$ there are corrections that are suppressed by $e^{ - \# N}$. At large $p$, there are a large number of non-perturbative corrections.

In the doubling scaling limit (\ref{riresult}) becomes %
\be
\hat{D}(j_R)=\frac{2}{p} \sum_{l=0}^{\infty}  \cos \left({\pi j_R(2 l+1)}\right)e^{-\frac{(2l+1)^2\pi^2}{4\lambda}}
\ee
which we reproduce from a bulk computation in the main text. 

\section{Inner product \la{innerProductM}}

In this Appendix, we derive the inner products between $\{\ket{n, \O\X, j},\ket{n, \O\O, j},\ket{n, \X\O, j},\ket{n, \X\X, j}\}$. This procedure is similar to Appendix C of \cite{superBerkooz}. First we should note that states with different number of chords or different charges are orthogonal to each other, so we only need to find the inner product within each sector. The idea is to express the inner product of two states with $2n$ (or $2n+1$) chords in terms of inner products of two states with $2n-1$ (or $2n$) chords by deleting the right-most chord from the configuration\footnote{See \cite{Lin:2022rbf} for analogous delete-a-chord relations in the $\nn = 0$ case.}. Then since we have the explicit formulas (\ref{jqrt}) and (\ref{jqrtb}), we can simplify the expressions and get recursion relations as follows. %

\begin{align}
\braket{n, \O\X, j}{n, \O\X, j} &= \mel{n-1,\O\O,j-1}{Q^{\text{del}}_\rt \bar{Q}^{\text{add}}_\rt }{n-1,\O\O,j-1} \\
    &= q^{j/2-3/4} \braket{n-1,\O\O,j-1} \\
\braket{n, \O\O, j}{n, \O\O, j} 
    &= \mel{n,\O\X,j+1}{\bar{Q}^{\text{del}}_\rt Q^{\text{add}}_\rt }{n, \O\X, j+1}+\mel{n,{\X\O},j+1}{\bar{Q}^{\text{del}}_\rt Q^{\text{add}}_\rt }{n,\O\X, j+1} \\ 
    &= \q^{-j/2-1/4}\braket{n,\O\X,j+1}+ \q^{n-j/2-1/4}\braket{n,{\X\O},j+1}{n, \O\X, j+1}\\ 
\braket{n, {\X\O}, j}{n, \O\X, j} &= \mel{n-1,\O\O,j-1}{Q^{\text{del}}_\rt \bar{Q}^{\text{add}}_\rt }{n-1,\O\O,j-1} \la{off-diag-overlap} \\
    &= -q^{(n-\hf)+j/2-1/4} \braket{n-1,\O\O,j-1} %
\end{align}
We have used a slightly schematic notation in the above; only one ``del'' term in $Q_R^\text{del}$ is used depending on the particular overlap in question. For example, in the last line \nref{off-diag-overlap}, the relevant delete term in $Q_R$ was $\text{del X}_L$. This is because all chord diagrams that contribute to the overlap \nref{off-diag-overlap} have a right-most chord which crosses over from right to left.
Defining
\begin{align}
    A_{n,j}&=\braket{n,\O\X,j}\\
    C_{n,j}&=\braket{n,{\X\O},j}{n,\O\X,j}\\
    a_{n,j}&=\braket{n,\O\O,j}
\end{align}
the recursion relations become
{
\begin{align}
    A_{n+1,j}&=q^{j/2-3/4}a_{n,j-1}\\
    a_{n,j-1}&=q^{-j/2+1/4}A_{n,j}+q^{n-j/2+1/4}C_{n,j}\\
    C_{n+1,j}&=-q^{n+j/2+1/4}a_{n,j-1}
\end{align}
}
Shifting the indices and eliminating the other variables gives a simple recursion relation for $A_{n,j}$:
\eqn{
A_{n+1,j}=q^{-1/2}(1-q^{2n})A_{n,j}%
}
The solution to the recursion relation (with the arbitrary initial condition $A_{0,j} = q^{-1/2}$) is
{%
\be
A_{n,j}=q^{-n/2}(q^2;q^2)_{n-1}.
\ee
}
We summarize the solutions below:
{%
\begin{align}
\braket{n,\O\X,j} &= q^{-n/2} (q^2; q^2)_{n-1}\\
\braket{n,{\X\O}}{n,\O\X} &= -q^{n/2} (q^2; q^2)_{n-1} \\
 \braket{n-1,\O\O,j-1} &= q^{-j/2+3/4 - n/2} (q^2; q^2)_{n-1}\\
 \braket{n,\O\O,j} &= q^{-j/2-1/4 - n/2} (q^2; q^2)_{n}
\end{align}
}
We may compare this to BBNR, their equation (4.37). We find that $\ket{n,\O\X,j}_\text{here} = q^{n/4} \ket{n,\O}_\text{BBNR}$. Hence we define
\begin{align}
    \alpha_{n,j} &= q^{n/4} \psi_{\X\O,j}(n) \la{alphaDef} \\
    \beta_{n,j} &= q^{n/4}  \psi_{\O\X,j}(n) \la{betaDef} %
\end{align}
This is defined so that $\psi_n \ket{n}_\text{here} = \alpha_n \ket{n}_\text{BBNR}$. Thus if we view $(\alpha,\beta)$ as the wavefunctions, the inner product agrees with the one found in BBNR.

\section{Derivation details \la{wvc}} 

In this appendix we prove a key identity that allows us to extract the number of ground states of the boundary SYK using purely bulk methods.

\subsection{Computing the norm \la{app:norm}}
In this section we will show that
\be
\la{normB}
\bra{\Psi,j}\ket{\Psi,j}
=\sum_{n=0}^\infty \frac{1}{(\q^2 ; \q^2)_n} \frac{\q^{3n}}{1-\q^{2n}}(\q^{-n}(a_n^2 + b_n^2) - 2 a_n b_n)=\frac{1}{(\q^{2j_R+1};\q^2)_\infty(\q^{-2j_R+1};\q^2)_\infty(\q^2;\q^2)_\infty} .
\ee
Here $j_R = -j/2$, and the coefficients $a_n$, $b_n$ are given by the q-Hermite polynomials
\begin{align}
a_n&=H_n\left(-\cosh(\lambda(j_R+\hf))|\q^2\right)=H_n(\cos\phi_a|\q^2) \quad\quad\phi_a=\pi-i\lambda(j_R+\frac{1}{2})\\
b_n&=H_n\left(-\cosh(\lambda(j_R-\hf ))|\q^2\right)=H_n(\cos\phi_b|\q^2)\quad\quad\phi_b=\pi-i\lambda(j_R-\frac{1}{2}) .
\label{eq:a_b_definition}
\end{align}
For $q<1$, this sum converges for $|j_R| < 1/2$, which is the expected range in which we find a normalizable (bound state) wavefunction.
These are related to the original wavefunctions $\alpha_n$, $\beta_n$ via
\be   
\alpha_n \equiv \frac{\q^{\frac{3n}{2}}}{(\q^2;\q^2)_n} a_n 
, \qquad \beta_n \equiv \frac{\q^{\frac{3n}{2}}}{(\q^2;\q^2)_n} b_n. \la{abdefAppendix}
\ee
As will be shown below, there are many cancelations in the above sum and the final result is in fact controlled by a single term as 
\be  
\bra{\Psi}\ket{\Psi}= \lim_{n\rightarrow \infty} 
\frac{\q^{n}}{(\q^2;\q^2)_{n}}
a_{n} b_{n} 
.
\ee
\\
\\
Let's start with some definitions. The q-Hermite polynomials can be defined in terms of q-Pochhammer symbols as 
\be  
H_n (\cos \phi | \q^2) = \sum_{k=0}^n \frac{(\q^2;\q^2)_n}{(\q^2;\q^2)_{n-k}(\q^2;\q^2)_k} e^{i(n-2k)\phi }  
.
\ee
To get the final result the following formula will also be useful
\be  
\frac{1}{(x;\q^2)_\infty} = \sum_{n=0}^\infty \frac{x^n}{(\q^2;\q^2)_n}
.
\ee
We denote $\tilde{\phi}_a = - i \lambda (j_R+\frac{1}{2})$, $\tilde{\phi}_b = - i \lambda (j_R-\frac{1}{2})$, in terms of which 
\begin{align}
a_n&=H_n(\cos\phi_a|\q^2) = (-1)^n H_n(\cos \tilde{\phi}_a|\q^2)
\\
b_n&=H_n(\cos\phi_b|\q^2) = (-1)^n H_n(\cos \tilde{\phi}_b|\q^2) .
\end{align}
To compute the norm it will be useful to find the relation between $a_n$ and $b_n$. Using the relation $\tilde{\phi}_a = \tilde{\phi}_b - i \lambda$ we have 
\be  
a_n 
= (-1)^n  \sum_{k=0}^n \frac{(\q^2;\q^2)_n}{(\q^2;\q^2)_{n-k}(\q^2;\q^2)_k} e^{i(n-2k)\tilde{\phi}_a } 
=  (-1)^n  \sum_{k=0}^n \frac{(\q^2;\q^2)_n}{(\q^2;\q^2)_{n-k}(\q^2;\q^2)_k} e^{i(n-2k)\tilde{\phi}_b} \q^{2k-n} ,
\ee
from which we can compute 
\begin{align}
\q^{-n} a_n - b_n  &= 
(-1)^n  \sum_{k=0}^n \frac{(\q^2;\q^2)_n}{(\q^2;\q^2)_{n-k}(\q^2;\q^2)_k} e^{i(n-2k)\tilde{\phi}_b} 
\frac{1 -\q^{2(n-k)}}{\q^{2(n-k)}}
\\
&= (-1)^n  \sum_{k=0}^{n-1} \frac{(\q^2;\q^2)_{n-1}(1-\q^{2n})}{(\q^2;\q^2)_{n-1-k}(\q^2;\q^2)_k} e^{i(n-2k)\tilde{\phi}_b} 
\frac{1}{\q^{2(n-k)}}
\\
&= (-1)^n  \sum_{k=0}^{n-1} \frac{(\q^2;\q^2)_{n-1}}{(\q^2;\q^2)_{n-1-k}(\q^2;\q^2)_k} e^{i(n-2k)\tilde{\phi}_a} 
\frac{1-\q^{2n}}{\q^{n}} 
\\
&= - \frac{1-\q^{2n}}{\q^{n}} e^{i\tilde{\phi}_a} a_{n-1} .
\end{align}
Therefore we find that 
\be  
b_n = \q^{-n} (a_n + e^{i\tilde{\phi}_a} 
(1-\q^{2n})a_{n-1}
) .
\label{eq:b_a_relation}
\ee
Inserting this relation into the expression for the norm we get
\be  
\bra{\Psi} \ket{\Psi} = \sum_{n=0}^\infty 
\frac{1}{(\q^2;\q^2)_n} 
\left( 
a_n^2 + (1-\q^{2n}) (2 a_{n}a_{n-1} e^{i\tilde{\phi}_a} + a_{n-1}^2 e^{2i\tilde{\phi}_a} ) 
\right) .
\ee
Now, using the recursion relation for $a_n$'s will result in many cancelations in the above sum. To analyze this carefully we rewrite the sum as
\footnote{We thank Sonja Klisch for suggesting this.}
\be  
\bra{\Psi} \ket{\Psi} = \lim_{n_{\max} \rightarrow \infty} \sum_{n=0}^{n_{\max}}
\frac{1}{(\q^2;\q^2)_n} 
\left( 
a_n^2 + (1-\q^{2n}) (2 a_{n}a_{n-1} e^{i\tilde{\phi}_a} + a_{n-1}^2 e^{2i\tilde{\phi}_a} ) 
\right) .
\ee
With this we now have 
\begin{align}
&\sum_{n=0}^{n_{\max}}
\frac{1}{(\q^2;\q^2)_n} 
\left( 
a_n^2 + (1-\q^{2n}) (2 a_{n}a_{n-1} e^{i\tilde{\phi}_a} + a_{n-1}^2 e^{2i\tilde{\phi}_a} ) 
\right)  \\
= & \sum_{n=0}^{n_{\max}}
\frac{a_n^2}{(\q^2;\q^2)_n} +
\sum_{n=0}^{n_{\max}}
\frac{(1-\q^{2n})}{(\q^2;\q^2)_n}
(2 a_{n}a_{n-1} e^{i\tilde{\phi}_a} + a_{n-1}^2 e^{2i\tilde{\phi}_a} ) 
\\
= & \frac{a_{n_{\max}}^2}{(\q^2;\q^2)_{n_{\max}}}  +\sum_{n=1}^{n_{\max}}
\frac{a_{n-1}^2}{(\q^2;\q^2)_{n-1}} +
\sum_{n=1}^{n_{\max}}
\frac{1}{(\q^2;\q^2)_{n-1}}
(2 a_{n}a_{n-1} e^{i\tilde{\phi}_a} + a_{n-1}^2 e^{2i\tilde{\phi}_a} )
\\
= & \frac{a_{n_{\max}}^2}{(\q^2;\q^2)_{n_{\max}}}  +
\sum_{n=0}^{n_{\max}-1}
\frac{e^{i\tilde{\phi}_a} a_n}{(\q^2;\q^2)_{n}}
( (e^{-i\tilde{\phi}_a}+e^{i\tilde{\phi}_a})a_n + 2 a_{n+1} ) .
\end{align}
Finally, applying the recursion relation
\be  
2\cos \phi_a a_n - a_{n+1} = -2\cos \tilde{\phi}_a a_n - a_{n+1}  =(1-\q^{2n}) a_{n-1} ,
\ee
to the last term gives 
\begin{align}
&\frac{a_{n_{\max}}^2}{(\q^2;\q^2)_{n_{\max}}}  +
\sum_{n=0}^{n_{\max}-1}
\frac{e^{i\tilde{\phi}_a} a_n}{(\q^2;\q^2)_{n}}
(a_{n+1} - (1-\q^{2n}) a_{n-1} ) 
\\= 
&\frac{a_{n_{\max}}}{(\q^2;\q^2)_{n_{\max}}} (a_{n_{\max}}+ e^{i\tilde{\phi}_a} (1-\q^{2n_{\max}}) a_{n_{\max}-1} )  = 
\frac{\q^{n_{\max}}}{(\q^2;\q^2)_{n_{\max}}}
a_{n_{\max}} b_{n_{\max}} 
.
\end{align}
We have therefore shown that the final answer for the norm can be written in a simple form as
\be  
\bra{\Psi}\ket{\Psi}= \lim_{n\rightarrow \infty} 
\frac{\q^{n}}{(\q^2;\q^2)_{n}}
a_{n} b_{n} 
.
\ee
The problem of finding the norm therefore reduced to finding the limit of the above product of q-Hermite polynomials. To take the limit we use the definitions and rewrite the resulting sums as
\begin{align}
\frac{\q^{n}}{(\q^2;\q^2)_{n}}
a_{n} b_{n}  &= \frac{\q^n}{(\q^2;\q^2)_{n}}  \sum_{k=0}^n \frac{(\q^2;\q^2)_n}{(\q^2;\q^2)_{n-k}(\q^2;\q^2)_k} e^{i(n-2k)\tilde{\phi}_a }   
\sum_{l=0}^n \frac{(\q^2;\q^2)_n}{(\q^2;\q^2)_{n-l}(\q^2;\q^2)_l} e^{i(n-2l)\tilde{\phi}_b }  
\\
&= 
\frac{1}{(\q^2;\q^2)_{n}}  \sum_{k=0}^n \frac{(\q^2;\q^2)_n}{(\q^2;\q^2)_{n-k}(\q^2;\q^2)_k} \q^{k(1+2j_R)}  
\sum_{l'=0}^n \frac{(\q^2;\q^2)_n}{(\q^2;\q^2)_{n-{l'}}(\q^2;\q^2)_{l'}}\q^{l'(1-2j_R)}
\\
& \xrightarrow[]{n \rightarrow \infty}  \,\,
\frac{1}{(\q^2;\q^2)_{\infty}}  \sum_{k=0}^\infty \frac{\q^{k(1+2j_R)}}{(\q^2;\q^2)_k}   
\sum_{l'=0}^\infty \frac{\q^{l'(1-2j_R)}}{(\q^2;\q^2)_{l'}} 
\\
& =
\frac{1}{(\q^{2j_R+1};\q^2)_\infty(\q^{-2j_R+1};\q^2)_\infty(\q^2;\q^2)_\infty}
,
\end{align}
where in taking the limit we used Tannery's theorem to take the limit under the sum.
\subsection{Computing the two point function \label{app:twoPt} } 
Using the techniques of the previous section we can also compute the two point function ($j=-2j_R$):
\be  
\bra{\Psi,j} \q^{2\Delta n} \ket{\Psi,j} = 
\sum_{n=0}^\infty \frac{\q^{2\Delta n}}{(\q^2 ; \q^2)_n} \frac{\q^{3n}}{1-\q^{2n}}(\q^{-n}(a_n^2 + b_n^2) - 2 a_n b_n) .
\ee
Here $a_n,b_n$ are defined in \nref{abdef}. Below we'll show that
\begin{align}
\la{2ptB2}
\bra{\Psi,j} \q^{2\Delta n} \ket{\Psi,j} 
&= \frac{(\q^{2+4\Delta} ; \q^2)_\infty}
{(\q^{2+2\Delta} ; \q^2)_\infty^2  (\q^{1\pm2j_R+2\Delta} ; \q^2)_\infty} .
\end{align}
The steps will closely follow the norm computation. We start with rewriting the infinite sum as 
\be  
\bra{\Psi,j_R}  \q^{2\Delta n} \ket{\Psi,j_R} = \lim_{n_m \rightarrow \infty} \sum_{n=0}^{n_m}
\frac{\q^{2\Delta n}}{(\q^2;\q^2)_n} 
\left( 
a_n^2 + (1-\q^{2n}) (2 a_{n}a_{n-1} e^{i\tilde{\phi}_a} + a_{n-1}^2 e^{2i\tilde{\phi}_a} ) 
\right) .
\ee
Splitting the sum as before and applying the recursion relation, the expression under the limit becomes
\begin{align}
=& \frac{\q^{(2\Delta+1 )n_{m}} a_{n_{m}} b_{n_{m}} }{(\q^2;\q^2)_{n_{m}}} 
+ (1-\q^{2\Delta}) \sum_{n=0}^{n_m -1} 
\frac{\q^{2\Delta n} }{(\q^2;\q^2)_{n}} a_n^2  
+ (1-\q^{2\Delta}) \sum_{n=0}^{n_m -2} 
\frac{\q^{(2\Delta+1 )n} }{(\q^2;\q^2)_{n}} e^{i\tilde{\phi}_a} a_{n+1} a_n \\
=&  (1-\q^{2\Delta}) \lb \sum_{n=0}^{n_m -1} 
\frac{\q^{2\Delta n} }{(\q^2;\q^2)_{n}} a_n^2  
+  \sum_{n=0}^{n_m -2} 
\frac{\q^{(2\Delta+1 )n} }{(\q^2;\q^2)_{n}} e^{i\tilde{\phi}_a} a_{n+1} a_n \rb 
\end{align}
Using now the relation in the last term
\be  
e^{i\tilde{\phi}_a} a_{n+1} = - \q^n b_n - e^{2i\tilde{\phi}_a} a_n ,
\ee
and taking the limit $n_{m} \rightarrow \infty$ the first term goes to zero (for $\Delta > 0$), and we are left with three convergent infinite sums over a product of q-Hermite polynomials which combine into
\be  
\bra{\Psi,j} \q^{2\Delta n} \ket{\Psi,j} = (1-\q^{2\Delta}) \sum_{n=0}^\infty 
\frac{\q^{2\Delta n}}{(\q^2;\q^2)_{n}}
\left( 
(1- e^{2i\tilde{\phi}_a}) \q^{2\Delta} a_n^2 -
\q^{n} \q^{2\Delta} a_n b_n 
\right)
.
\ee
To both sums above we can now apply the relation 
\be   
\sum_{n=0}^\infty \frac{ t^n H_n(\cos \phi |\q^2)H_n(\cos \phi' |\q^2)}{(\q^2;\q^2)_n} 
= \frac{(t^2,\q^2)_\infty}{(t  e^{i(\pm \phi \pm \phi')};\q^2)_\infty}
,
\label{eq:qHermite_orthogonality}
\ee
which after rewriting leads us to the final answer.

\subsection{BPS two
point function}
\label{app:BPStwo_pt_function}
Similarly we can compute a two point function of a charged BPS operator with the $U(1)_{\mathrm{R}}$ charge $J_R=-J_L=2\Delta$. Here we will show that ($j=-2j_R$, $j'=-2j'_R$)
\be  
\bra{\Psi,j'}\q^{2\Delta n} (\tilde{\beta}^\dag)^{4\Delta} 
q^{-\Delta j}
\ket{\Psi,j}
= \delta_{j'_R,j_R-2\Delta}\,
q^{2 \Delta j_R}
\,
\frac{1}
{(\q^{2} ; \q^2)_\infty  
(\q^{1+2j_R} ; \q^2)_\infty
(\q^{1-2j'_R} ; \q^2)_\infty
}
.\label{eq:2pt_function_charged}
\ee
To derive this, it is convenient to make use of the orthonormal basis 
\begin{align}
\ket{n_\pm , j} = \frac{\q^{n/4}}{\sqrt{2 (\q^2 ,\q^2)_{n-1} (1\mp \q^n) }} (\ket{n,\O\X ,j} \pm \ket{n,\X\O , j}) .
\end{align}
We have then 
\begin{align}
\bra{\Psi,j'}\q^{2\Delta n} (\tilde{\beta}^\dag)^{4\Delta} \ket{\Psi,j} &= 
\sum_{n}
\bra{\Psi,j'}\ket{n_\pm , j+4\Delta} \q^{2\Delta n} \bra{n_\pm ,j} \ket{\Psi,j} 
\\
& = \delta_{j',j+4\Delta} \sum_{n} \bra{\Psi, j'} \ket{n_\pm , j'} \q^{2\Delta n} \bra{n_\pm ,j} \ket{\Psi, j} .
\end{align}
The rest of the computation follows closely the computation of the uncharged case. We need to evaluate the sum 
\be  
\sum_{n} \bra{\Psi, j'} \ket{n_\pm , j'} \q^{2\Delta n} \bra{n_\pm ,j} \ket{\Psi, {j}}
= \sum_n \frac{\q^{2\Delta n + 2n} (\q^2 ,\q^2)_{n-1} }{(\q^2 ,\q^2)_n ^2} 
(b_n (b'_n - \q^n a'_n) + a_n (a'_n-b'_n\q^n)) 
,
\label{eq:2pt_charged_sum}
\ee
with $a'_n,$ $b'_n$, defined as in \eqref{eq:a_b_definition} but with $j_R \rightarrow j'_R$. 
Below, analogously to $\tilde{\phi}_{a/b}$ we'll denote $\tilde{\phi}_{A/B} \equiv -i\lambda(j'_R \pm\frac{1}{2})$. Expressing now the above sum fully in terms of $a_n^j , a_n^{j'}$ with the relation \eqref{eq:b_a_relation} cancels the factor $(1-\q^{2n})$. Collecting the resulting terms one can rewrite the right hand side of \eqref{eq:2pt_charged_sum} as 
\be  
\sum_n \frac{\q^{2\Delta n}}{(\q^2 , \q^2)_n} \left( 
(1-\q^{2\Delta}e^{\i(\tilde{\phi}_a + \tilde{\phi}_A)})a_n^j a_n^{j'} 
- \q^{2\Delta} \q^n (e^{\i(\tilde{\phi}_a - \tilde{\phi}_A)} a_n^j b_n^{j'}
+ e^{\i(\tilde{\phi}_A - \tilde{\phi}_a)} a_n^{j'} b_n^{j}
) 
\right) 
.
\ee
Applying now the \eqref{eq:qHermite_orthogonality} to each of the terms above yields \eqref{eq:2pt_function_charged}.

\section{Matching with the Schwarzian description \la{schMatch} }

Let's see how we can reproduce the super Schwarzian ground state wavefunction using $\alpha_n$ and $\beta_n$. 
We will use the notation of \cite{LongPaper}.
We can take the ansatz for the ground state $\ket{Z_{j}}$ in the form 
\be  
\ket{Z_{j}} = \left( 
e^{i a (j_R+\frac{1}{2})} h_r(\ell) \bar{\psi}_r 
+ 
e^{i a(j_R-\frac{1}{2})} h_l (\ell) \bar{\psi}_l
\right) \ket{\frac{1}{2},\frac{1}{2}} ,
\ee
Imposing that the ground state is annihilated by the supercharges
\be  
\bar{Q}_r = \bar{\psi}_r (i \partial_\ell - \frac{1}{2} \partial_a) + e^{-\frac{\ell}{2}-ia} \bar{\psi}_l , 
\qquad 
\bar{Q}_l = \bar{\psi}_l (i \partial_\ell + \frac{1}{2} \partial_a) - e^{-\frac{\ell}{2}+ia} \bar{\psi}_r ,
\ee
leads to equations for $h_l(\ell),h_r(\ell)$:
\begin{align}
i h_r'(\ell) + \frac{i}{2}\left(j_R + \frac{1}{2} \right) h_r(\ell) + e^{-\frac{\ell}{2}} h_l(\ell) &= 0 ,
\\ 
i h_l'(\ell) - \frac{i}{2}\left(j_R - \frac{1}{2} \right) h_l(\ell) - e^{-\frac{\ell}{2}} h_r(\ell) &= 0 .
\end{align}
From these equations we find the unique wavefunctions (up to an overall multiplicative constant)
\begin{align}
h_l (\ell) &= i e^{-\frac{\ell}{2}}  K_{\frac{1}{2}+j_R} (2 e^{-\frac{\ell}{2}})
\\ 
h_r (\ell) &=  - e^{-\frac{\ell}{2}} K_{\frac{1}{2}-j_R} (2 e^{-\frac{\ell}{2}})
.
\end{align}
We will now analyze the $\lambda \rightarrow 0$ limit of the differential equation that defines $\alpha_n$ and $\beta_n$ and see how they reproduce the above super-Schwarzian wavefunctions. 
\\ 
\\ 
To normalize these wavefunctions, we will need the inner product. This takes the form: 
\be  
\braket{Z_j} = \int d \ell (h_r^* h_r + h_l^* h_l) 
=  \int d \ell \, e^{-\ell} \left( 
K_{\frac{1}{2}-j_R} (2 e^{-\frac{\ell}{2}})^2 
+ 
K_{\frac{1}{2}+j_R} (2 e^{-\frac{\ell}{2}})^2
\right) = \frac{\pi}{4 \cos(\pi j_R)} 
.
\ee

Recall that the solutions to $\alpha_n$ and $\beta_n$ oscillates 
\begin{align}
\alpha_n&=(-1)^n\frac{\q^{\frac{3n}{2}}}{(\q^2;\q^2)_n}H_n\left[\cosh\left(\lambda(j_R+\tfrac{1}{2})\right)\Bigg|\q^2\right]\\
\beta_n&=(-1)^n\frac{\q^{\frac{3n}{2}}}{(\q^2;\q^2)_n}H_n\left[\cosh\left(\lambda(-j_R+\tfrac{1}{2})\right)\Bigg|\q^2\right]
\end{align}
In order to take $\alpha_n$ and $\beta_n$ to the continuous limit, we need to separate out a rapidly oscillation component, i.e. 
\be
\alpha_n=(-1)^n\tilde{\alpha}_n\quad\quad\beta_n=(-1)^n\tilde{\beta}_n
\ee
Then the recursion relations become
\be 
-q^n \tilde{\alpha}_n + \tilde{\beta}_n - q^{j_R+1} \tilde{\beta}_{n-1} =0 , 
\qquad 
\tilde{\alpha}_n - q^n \tilde{\beta}_n - q^{1-j_R} \tilde{\alpha}_{n-1} =0.
\ee
To find the leading terms in $\lambda \rightarrow 0$ limit we redefine $n = \ell/2\lambda$ and take 
\be  
\tilde{\alpha}_{n\pm 1} = \tilde{\alpha} (\ell) \pm 2\lambda \tilde{\alpha}'(\ell) + O(\lambda^2) ,\la{diffApprox}
\ee
and similarly for $\tilde{\beta}_n$. From this we obtain two first order differential equations 
\be  
- e^{-\frac{\ell}{2}} \tilde{\alpha}(\ell) + 
\lambda(j_R+1) \tilde{\beta}(\ell) + 2\lambda \tilde{\beta}'(\ell) = 0 , \qquad 
- e^{-\frac{\ell}{2}} \tilde{\beta}(\ell) + 
\lambda(1-j_R) \tilde{\alpha}(\ell) + 2\lambda \tilde{\alpha}'(\ell) = 0 .
\ee
or introducing the renormalized length $\tilde{\ell} \equiv \ell + 2\log(2\lambda)$,
\begin{align}
- e^{-\frac{\tilde{\ell}}{2}} \tilde{\alpha} (\tilde{\ell})  - \frac{1}{2}(1+j_R) \tilde{\beta}(\tilde{\ell}) - \tilde{\beta}'(\tilde{\ell}) &= 0 , \la{d16}
\\ 
e^{-\frac{\tilde{\ell}}{2}} \tilde{\beta} (\tilde{\ell}) - \frac{1}{2}(1-j_R) \tilde{\alpha} (\tilde{\ell}) 
- \tilde{\alpha}' (\tilde{\ell}) \la{d17}
&=0 .
\end{align}

These equations are now solved by 
\begin{align}
\tilde{\alpha} (\ell) &= \sqrt{\mathcal{N}}
 e^{-\frac{3}{4}\tilde\ell}  K_{\frac{1}{2}+j_R} \left( 
{e^{-\tilde\ell/2}}
\right)
, \\ 
\tilde{\beta} (\ell) &=  \sqrt{\mathcal{N}}
 e^{-\frac{3}{4}\tilde\ell } K_{\frac{1}{2}-j_R} \left( 
{e^{-\tilde\ell/2}}
\right)
.
\end{align}
Rescaling 
\begin{align} 
\tilde{\beta} (\tilde{\ell}) \equiv 
e^{-\frac{\tilde{\ell}}{4}}
h_r(\tilde{\ell}) ,  
\qquad
\tilde{\alpha} (\tilde{\ell}) \equiv 
i e^{-\frac{\tilde{\ell}}{4}} 
h_l(\tilde{\ell}) ,
\end{align}
the above equations \nref{d16}, \nref{d17} reduce to 
\begin{align}
i h_r'(\tilde{\ell}) + \frac{i}{2}\left(j_R + \frac{1}{2} \right) h_r(\tilde{\ell}) + e^{-\frac{\tilde{\ell}}{2}} h_l(\tilde{\ell}) &= 0 ,
\\ 
i h_l'(\tilde{\ell}) - \frac{i}{2}\left(j_R - \frac{1}{2} \right) h_l(\tilde{\ell}) - e^{-\frac{\tilde{\ell}}{2}} h_r(\tilde{\ell}) &= 0 ,
\end{align}
which are exactly the equations we obtained from super-Schwarzian theory above. 
\\
To evaluate the leading part of the norm we want to rewrite the expression in terms of $\tilde{\ell}_n = 2\lambda n + 2 \log(2\lambda)$
\begin{align}
\braket{\Psi, j} &= 
\sum_n (\q^2; \q^2)_{n-1}  \lb \q^{-n} \lp \alpha^2_n + \beta^2_n \rp - 2\alpha_n \beta_n \rb  
\\ 
&\simeq  \mathcal{N} 
\sum_n   
\frac{(\q^2; \q^2)_{n-1} }{2\lambda} e^{-\tilde{\ell}_n} \lb
K_{\frac{1}{2} +j_R}(2 e^{-\frac{\tilde{\ell}_n}{2}})^2 
+ K_{\frac{1}{2} -j_R}(2 e^{-\frac{\tilde{\ell}_n}{2}})^2 
- 4\lambda e^{-\frac{\tilde{\ell}_n}{2}}
K_{\frac{1}{2} +j_R}
K_{\frac{1}{2} -j_R}
\rb  
\notag \\ 
&\simeq   \mathcal{N} 
\sum_n   
\frac{(\q^2; \q^2)_{n-1} }{2\lambda}  
e^{-\tilde{\ell}_n} \lb
K_{\frac{1}{2} +j_R}(2 e^{-\frac{\tilde{\ell}_n}{2}})^2 
+ K_{\frac{1}{2} -j_R}(2 e^{-\frac{\tilde{\ell}_n}{2}})^2 
\rb  \notag \\
&\simeq   \mathcal{N} 
\sum_n   
\frac{(\q^2; \q^2)_\infty }{2\lambda}  
e^{-\tilde{\ell}_n} \lb
K_{\frac{1}{2} +j_R}(2 e^{-\frac{\tilde{\ell}_n}{2}})^2 
+ K_{\frac{1}{2} -j_R}(2 e^{-\frac{\tilde{\ell}_n}{2}})^2 
\rb  \notag \\
&\simeq   \mathcal{N} \frac{(\q^2; \q^2)_\infty }{(2\lambda)^2} 
\int d\tilde{\ell}   
e^{-\tilde{\ell}_n} \lb
K_{\frac{1}{2} +j_R}(2 e^{-\frac{\tilde{\ell}_n}{2}})^2 
+ K_{\frac{1}{2} -j_R}(2 e^{-\frac{\tilde{\ell}_n}{2}})^2 
\rb  \notag \\
&\simeq   \mathcal{N} \frac{(\q^2; \q^2)_\infty }{(2\lambda)^2} \frac{\pi}{4 \cos(\pi j_R)} ,
\end{align}
where 
we should now fix the normalization $\mathcal{N}$ to match the original answer at small $\lambda$. This will fix the normalization of the solutions to the differential equation. This means we have 
\be  
\mathcal{N} \frac{(\q^2; \q^2)_\infty }{(2\lambda)^2} \frac{\pi}{4 \cos(\pi j_R)}
= \frac{1}{(\q^{1 \pm 2j_R};\q^2)_\infty (\q^2;\q^2)_\infty} .
\ee
Because 
\be  
\frac{1}{(\q^{1 \pm 2j_R};\q^2)_\infty (\q^2;\q^2)_\infty} = \frac{\Gamma_{q^2}(\frac{1}{2} +j_R)\Gamma_{q^2}(\frac{1}{2} -j_R)}{(q^2 ; q^2)_\infty^3 (1-q^2)} ,
\ee
at small $\lambda$ we can drop the $j_R$-dependence in the ratio appearing above $\Gamma_{q^2}(\frac{1}{2}  \pm j_R)  \simeq \frac{\pi}{\cos(\pi j_R)}$
\be  
\frac{8 \lambda  \cos (\pi j_R) }{(\q^{1 \pm 2j_R};\q^2)_\infty (\q^2;\q^2)_\infty \pi  } \simeq \frac{8\lambda}{(\q^2;\q^2)_\infty^3 (1-q^2)} . 
\ee
We therefore find that the normalization is given by
\begin{align}
\mathcal{N} 
=\frac{16\lambda^2}{(q^2;q^2)_\infty^4(1-q^2)} 
= \frac{8\lambda}{(q^2;q^2)_\infty^4} .
\end{align}
We also checked numerically that with this normalization factor the approximated wavefunctions $\tilde{\alpha},\tilde{\beta}$
matched the original wavefunctions $\alpha_n ,\beta_n$ at small $\lambda$.%

\section{Subalgebra of the superchord algebra \la{app:subalg} }
The $\chi_{ij}$  and $\phi_{ij}$ generators satisfy the following relations, which we now list.
\begin{align}
 [\chi_{LL}, \chi_{RR}] &= q^{\hf (n_\tot -1)} (\chi_{LR} -\chi_{RL}) \\
 [\chi_{RL}, \chi_{LR}] &= q^{\hf (n_\X - n_\O -1)} (\chi_{RR} -\chi_{LL}) \\
\chi_{LL}\chi_{LR} &= \chi_{LR} \chi_{RR} = q^{\hf (n_\X -n_\O - 1) } \chi_{LR} \\
 \chi_{RR}\chi_{RL} &= \chi_{RL} \chi_{LL} = q^{\hf (n_\X -n_\O - 1) } \chi_{RL} \\
 \chi_{LL}\chi_{RL} &= \chi_{LR} \chi_{LL} = q^{\hf (n_\tot - 1) } \chi_{LL} \\
 \chi_{RR}\chi_{RL} &= \chi_{RL} \chi_{RR} = q^{\hf (n_\tot - 1) } \chi_{RR} 
\end{align}
Next we consider relations amongst the $\phi_{ij}$:
\begin{align}
 [\phi_{LL}, \phi_{RR}] &= q^{\hf (n_\tot -1)} (\phi_{RL} -\phi_{LR}) \\
 [\phi_{RL}, \phi_{LR}] &= q^{\hf (n_\O - n_\X -1)} (\phi_{RR} -\phi_{LL}) \\
 \phi_{LL}\phi_{LR} &= \phi_{LR} \phi_{RR} = q^{\hf (n_\O -n_\X - 1) } \phi_{LR} \\
 \phi_{RR}\phi_{RL} &= \phi_{RL} \phi_{LL} = q^{\hf (n_\O -n_\X - 1) } \phi_{RL} \\
 \phi_{LR}\phi_{LL} &= \phi_{LL} \phi_{RL} = -q^{\hf (n_\tot - 1) } \phi_{LL} \\
 \phi_{RR}\phi_{LR} &= \phi_{RL} \phi_{RR} = -q^{\hf (n_\tot - 1) } \phi_{RR} 
\end{align}

Finally, we consider relations that involve both $\phi$ and $\chi$:
  \begin{align}
      [\phi_{RR}, \chi_{LL}] &= [\phi_{LL}, \chi_{RR}] = [\phi_{LR}, \chi_{RL}] = [\phi_{RL}, \chi_{LR}] =  0 %
  \end{align}

We found a special 1-parameter family of operators
\eqn{
\mathcal{C}_{\chi} &= \left\{\chi_{LL} - a_\chi q^{\hf(n_\X - n_\O - 1)}, \; \chi_{RR} - a_\chi q^{\hf(n_\X - n_\O - 1)} \right\} +  (2a_\chi-1) q^{\hf (n_\tot - 1)} (\chi_{LR} + \chi_{RL}) \\
\mathcal{C}_{\phi} &= \left\{\phi_{LL} - a_\phi q^{\hf(n_\O - n_\X - 1)}, \; \phi_{RR} - a_\phi q^{\hf(n_\O - n_\X - 1)} \right\} -  (2a_\phi-1) q^{\hf (n_\tot - 1)} (\phi_{LR} + \phi_{RL}) 
}
Here $C_\chi$ is the Casimir of the sub-algebra generated by $\chi$ and similarly $C_\phi$ is the Casimir of the sub-algebra generated by $\phi$. Presumably by combining these operators one can build a Casimir for the algebra generated by both $\chi_{ij}$ and $\phi_{ij}$, which could then be promoted to the Casimir of the full superchord algebra.

\section{The \texorpdfstring{$\mathcal{N}=1$}{N=1} SYK model \label{app:n1}} 
In this section, we study the double-scaled $\nn = 1$ SUSY SYK \cite{Fu:2016vas} in the chord formalism. 
This model is somewhat simpler than the $\nn = 2$ model since it has no additional global symmetries. However, it does not have exact BPS ground states.

The basic idea is to study moments of the supercharge $Q$ instead of moments of $H$. 
Each $Q$ has a similar form to the Hamiltonian in the $\nn = 0$ model, except with $q \to -q$. We take $r \to \pm r$ depending on whether the matter operator is fermion even (+) or odd ($-$). We will focus on the case where $r$ is even.
 
\subsection{The \texorpdfstring{$\mathcal{N}=1$}{N=1} chord algebra}
\def\lr{{\lt/\rt}}
\def\rl{{\rt/\lt}}
\def\b{\mathfrak{b}}

This gives
\begin{equation}
\begin{split}
    Q_\lt  &=  \cre_\lt+  \alpha_\lt [n_L]_{-q} + \alpha_\rt \, r (-q)^{n_\lt} [n_R]_{-q}   ,\\
    Q_\rt  &= (-1)^{F}  \lp \cre_\rt +  \alpha_\rt [n_R]_{-q}  + \alpha_\lt r (-q)^{n_\rt } [n_L]_{-q} \rp   ,\\
\end{split}
\end{equation}
Following the notation of \cite{newpaper}, we have
\begin{align}
    Q_\lt &= \b_L + \b_L^\dagger, \\
    Q_\rt &= \b_R + \b_R^\dagger 
\end{align}
Then we have
\begin{align}
&\{\b_L, \b_R \} = \{\b_L^\dagger, \b_R^\dagger\} = 0\\
&[\nb, \b^\dagger_{\lr} ] = \b^\dagger_{\lr}, \quad [\nb,\b_{\lr}] = -\b_{\lr}\\
&\{\b_R, \b_L^\dagger \} = - \{\b_L, \b_R^\dagger\} =  q^\nb \\
&\{ \b_L , \b_L^\dag \}_q = - \{ \b_R , \b_R^\dag \}_q = 1\\
& \{ (-1)^F, \b_{L/R} \} = \{ (-1)^F, \b^\dag_{L/R} \} = [(-1)^F, \nb ] = 0
\end{align}
This algebra comes with a coproduct $D$ which is a co-associative, bilinear map $\mathcal{A} \to \mathcal{A} \otimes \mathcal{A}$:
\begin{align}
D(\mathfrak{b}_L^{\dagger}) & =\mathfrak{b}_L^{\dagger} \otimes 1, \quad D(\mathfrak{b}_R^{\dagger})= (-1)^F \otimes \mathfrak{b}_R^{\dagger}, \\
D\left(\mathfrak{b}_L\right) & =\mathfrak{b}_L \otimes 1 + q^{\Delta} q^{\bar{n}} (-1)^F \otimes \mathfrak{b}_L, \\
D\left(\mathfrak{b}_R\right) & = (-1)^F \otimes \mathfrak{b}_R + q^{\Delta} \mathfrak{b}_R \otimes q^{\bar{n}}, \\
D(\bar{n}) & =\bar{n} \otimes 1+1 \otimes \bar{n}+\Delta(1 \otimes 1)\\
D((-1)^F) &= (-1)^F \otimes (-1)^F
\end{align}
One can verify that the coproduct is an algebra homomorphism. A sample calculation is:
\begin{align}
     D(\{\b_L,\b_L^\dag \}_q ) &= \{ D(\b_L), D(\b_L^\dag) \} _q \\ %
    & = 1 + q^\Delta \{ q^\nb (-1)^F , \b^\dag \}_q \otimes \b_L \\
    &= 1 
\end{align}
The last line relies on $q^\nb (-1)^F \b_L^\dag  =-  \b_L^\dag q^{\nb+1} (-1)^F$. %
Another sample calculation is
\begin{align}
D( \{ \b_L^\dag, \b_R^\dag \} ) = \{ \b_L^\dag , (-1)^F \} \otimes \b_R^\dag = 0. %
\end{align}
\def\ta{\tilde{\alpha}}
\def\tad{\tilde{\alpha}^\dagger}
\def\ta{b}
\def\tad{b^\dagger}

Note that since $Q$ always changes the total number of chords by 1, the operator $Q^2$ always changes the total number by 0 or 2. Therefore, $T=Q^2$ decomposes into an even and odd chord number sector. This becomes the fermionic and bosonic states in the 2-sided Hilbert space.

We may also write the algebra in terms of just the $\b^\dagger$ and $Q$: %
\begin{equation}
\begin{split}
\label{qdefsuper}
  \{ Q_\lt, Q_\rt\} &=0\\
 [Q_{\lt/\rt},\nb]  &=  Q_\lr - 2 \b_\lr^\dag \\
  	 \{ \b_\lt^\dag, \b_\rt^\dag\} &=0\\
  	  [\nb, \b_\lr^\dag] &= \b_\lr^\dag \\
  	\{Q_\lt , \b_\lt^\dag \}_q &=1+ (1+q)(\b_\lt^\dag)^2  \\
  	 \{Q_\rt , \b_\rt^\dag \}_q &= -1+(1+q)(\b_\rt^\dag)^2  \\
		\{Q_{\lt}, \b_{\rt}^\dag\} &= -q^{\nb} \\
		\{Q_{\rt}, \b_{\lt}^\dag\} &= q^{\nb} \\
\{Q_\lr ,  	(-1)^F\} &=  [\nb , (-1)^F] = 0
\end{split}
\end{equation}
This is formally related to the $\caln = 0$ bosonic algebra \cite{newpaper} by taking 
\begin{equation}
\begin{split}
\label{continuation}
&  \mathsf{h}_\lt \to Q_\lt, \quad \mathsf{h}_\rt \to (-1)^n Q_\rt\\
&\mathfrak{a}^\dag_\lt \to \b_\lt^\dag, \quad \mathfrak{a}_\rt^\dag \to (-1)^n\b_\rt^\dag \\
 & q\to -q 
\end{split}
\end{equation}

An interesting subalgebra is generated by
\begin{equation}
\begin{split}
\label{neq1subalg}
  f_{ij} =\b_i^\dag \b_j  , \quad [f_{ij}, \nb] = 0 \\ 
\end{split}
\end{equation}
These elements satisfy
\def\lr{{\lt\rt}}
\def\rl{{\rt\lt}}
\def\ll{{\lt\lt}}
\def\rr{{\rt\rt}}
\begin{equation}
\begin{split}
\label{neq1subalg2}
  [f_{\rt\rt},f_{\lt\lt}] &= q^{\nb-1} (f_{\lt\rt} + f_{\rt\lt})\\
  [f_{\rt\lt},f_{\lt\rt}] &=  f_{\lt\lt} + f_{\rt\rt}\\
  \{f_{\ll}, f_{\lr} \}_q  &= f_{\lr}+q^{\nb} f_\ll \\
  -\{f_{\rr}, f_{\rl} \}_q  &= f_{\rl}+q^{\nb} f_\rr \\
  \{f_\rl, f_\ll\}_q &= f_\rl -q^{\nb} f_\ll\\
  -\{f_\lr, f_\rr\}_q &= f_\lr - q^{\nb} f_\rr.
\end{split}
\end{equation}

It would be interesting to characterize the symmetries in the triple scaling limit. One would like to make contact with the superalgebra $\mathfrak{osp}(1|2)$, which is the symmetry algebra of the bulk $\mathcal{N}=1$ JT theory. %

Using some inspiration from the $\mathcal{N} = 0$ case, we were able to guess the form of a Casimir:
\eqn{ \mathcal{C} = (-1)^F \left( q^{n-1} \{ 1-(1+q) f_{LL}, 1+ (1+q) f_{RR} \} + (1-q^2) f_{LR} - (1-q^2) f_{RL} - 2 q^n \right) }
This Casimir commutes with $n$ and also both supercharges, so it commutes with the entire algebra. On 1-particle states it gives
\eqn{  \mathcal{C}  \ket{ \text{1 particle}} = 2  \left(\frac{q}{r}-r\right ) }

\subsection{Fake superspace? \la{Liouville}}
\def\cg{\mathcal{G}}
\def\frg{\mathfrak{g}}
\def\sgn{\mathop{\mathrm{sgn}}}
\def\ss{{\psi\psi}}
The finite $p$ equations of motion for the $\caln = 1$ model is \cite{Fu:2016vas} 
\begin{equation}
D_{\theta_{1}} \mathcal{G}\left(\tau_{1}, \theta_{1} ; \tau_{3}, \theta_{3}\right)+J \int d \tau_{2} d \theta_{2} \mathcal{G}\left(\tau_{1}, \theta_{1} ; \tau_{2}, \theta_{2}\right)\left( \mathcal{G}\left(\tau_{2}, \theta_{2} ; \tau_{3}, \theta_{3}\right)^{p-1}\right)=\left(\theta_{1}-\theta_{3}\right) \delta\left(\tau_{1}-\tau_{3}\right)
\end{equation}
Now to take the large $p$ limit, we assume that the superfield exponentiates:
\eqn{\cg(\tau_1, \theta_1; \tau_2, \theta_2) &= \cg_0 e^{\frg/p},\\
\cg_0(\tau_1,\theta_1; \tau_2 \theta_2) &= 
 \hf \sgn(\tau_1 - \tau_2 - \theta_1 \theta_2) }
Notice that $\cg_0$ solves the free equations of motion:
\eqn{D_{\theta_1} \cg_0 = (\theta_1 -\theta_2) \delta(\tau_{12})}
Then we can get the large $p$ equations of motion by differentiating and plugging in the ansatz
\eqn{D_{\theta_2} D_{\theta_{1}} \frg - \cj e^\frg = 0 \la{superLiouville} }
This is solved at zero temperature by
\eqn{e^{\frg} = \frac{1}{\mathcal{J} (\tau_{1} - \tau_2 - \theta_1 \theta_2) + 1  } }
If we expand $\frg = g_{\psi\psi} +\theta_1 \theta_2 g_{bb}$, we may check that the component equations of motion \nref{superLiouville} are satisfied:
\eqn{
0&= g_{bb} -  \cj e^{g_{\psi \psi}}\\
0 &= \theta_1 \theta_2 \lb \pd_1 \pd_2 g_{\psi \psi} - \cj e^{g_\ss}  g_{bb}    \rb   }
The resulting expression looks like a conformally invariant function in superspace, with a fake region given by $-\cj\inv < \tau_1 -\tau_2 <0$.

\bibliography{bibChords.bib}
\end{document}